\documentclass[
    aps,
    twocolumn,
    nopacs,
    preprintnumbers,
    noeprint,
]{revtex4-2}

\usepackage[T1]{fontenc}

\usepackage{graphicx}
\usepackage{amsfonts}
\usepackage{amsmath}
\usepackage{amssymb}
\usepackage{glossaries}
\usepackage{siunitx}
\usepackage{xcolor}
\usepackage{float}
\usepackage{layouts}

\usepackage{xr}

\usepackage{hyperref}

\hypersetup{
    colorlinks=true,
    linkcolor=blue,
    citecolor=teal,
    filecolor=magenta,
    urlcolor=cyan,
    pdftitle={A new architecture for high-quality gate-defined structures in bilayer graphene},
    pdfpagemode=UseNone
}

\makeatletter

\newcommand*{\addFileDependency}[1]{%
    \typeout{(#1)}%
    \@addtofilelist{#1}%
    \IfFileExists{#1}{}{%
        \typeout{No file #1.}%
    }%
}

\newcommand*{\myexternaldocument}[2][]{%
    \externaldocument[#1]{#2}%
    \addFileDependency{#2.tex}%
    \addFileDependency{#2.aux}%
}

\makeatother

\myexternaldocument[SI-]{supplementary}

\newcommand{\subtxt}[2]{\ensuremath{#1_{\mathrm{#2}}}}

\newacronym{blg}{BLG}{bilayer graphene}
\newacronym{fet}{FET}{field-effect transistor}
\newacronym{hbn}{h-BN}{hexagonal boron nitride}
\newacronym{rie}{RIE}{reactive-ion etching}
\newacronym{sdh}{SdH}{Shubnikov–de Haas}
\newacronym{kpfm}{KPFM}{Kelvin probe force microscopy}

\DeclareSIUnit{\torr}{Torr}
\DeclareSIUnit{\sccm}{sccm}

\begin{document}

\title{Isolating the natural edges of bilayer graphene in gate-defined mesoscopic devices}

\author{Francesco Blanda$^1$}
\author{Grazia Raciti$^1$}
\author{Thilo Glatzel$^1$}
\author{Aurin Strathmann$^1$}
\author{Fabrizio Volante$^1$}

\author{Kenji Watanabe$^2$}
\author{Takashi Taniguchi$^2$}

\author{Ilaria Zardo$^1$}

\author{Thomas Ihn$^3$}
\author{Klaus Ensslin$^3$}

\author{Andrea Hofmann$^{1,*}$}

\begin{abstract} 
\begin{center}
\textsuperscript{1} University of Basel, Klingelbergstrasse 82, 4056 Basel, Switzerland \\
\textsuperscript{2} Advanced Materials Laboratory, National Institute for Materials Science, 1-1 Namiki, Tsukuba 3050044, Japan \\
\textsuperscript{3} Solid State Physics Laboratory, ETH Zürich, 8093 Zürich, Switzerland \\
\textsuperscript{4} Swiss Nanoscience Institute, Klingelbergstrasse 82, 4056 Basel, Switzerland\\
\textsuperscript{*} Corresponding author. Email: andrea.hofmann@unibas.ch\\
\vspace{1em}

\date{\today}
\end{center}
We introduce a graphite-gated architecture for bilayer graphene devices in which the active device is completely isolated from the natural graphene edges. Using a single patterned graphite-gate layer, we realize a fully electrostatically defined Hall-bar.
Longitudinal and Hall measurements reveal mesoscopic transport features, including Hall-effect quenching and magnetoresistance peaks associated with boundary scattering. The dependence of the mesoscopic features on the carrier-density shows that the effective channel width increases with the Fermi level and the electrostatic confinement at the gate-defined boundaries, and indicates that the carriers scatter at the electrostatic boundary. Raman spectroscopy and Kelvin probe force microscopy suggest that this boundary is disordered due to the used fabrication methods. 
Comparably, the quantum mobility in a field-effect transistor fabricated with the same architecture is not limited by boundary scattering and the visibility of quantum oscillations down to \SI{4}{\milli\tesla} suggests a record value of \SI{2.5e6}{\centi\meter^2/\volt\second}.

\end{abstract}


\maketitle 

\section{Introduction}

In the field of Van-der Waals materials, device reproducibility and yield impose major limitations to the exploration of the underlying physics and to their technological relevance. The complex assembly and fabrication schemes give rise to an immense number of parameters that have to be controlled at the same time. In \gls{blg} devices, the mastery of this parameter space is, to some extent, reflected by the opening of a uniform band-gap under the application of an electric field perpendicular to the \gls{blg} sheet. This gap is theoretically expected \cite{mccann_asymmetry_2006} and routinely achieved in experiments \cite{castro_biased_2007, zhang_direct_2009}, but its cleanliness as measured by the in-gap resistance is of varying quality and strongly depends on the fabrication details \cite{icking_transport_2022}. In the presence of a clean band-gap, the resistance increases beyond the Teraohm regime when the Fermi energy is tuned into the gap. In a device, conducting regions and insulating regions can be separated with patterned gate electrodes which tune the Fermi level into the band or the band gap, respectively. This allows to electrostatically define nanostructured devices. These structures have replaced etch-defined \gls{blg} architectures \cite{bischoff_importance_2016}, where the device was defined by etching away the surrounding material and allowed for measurements of clean quantum point contacts \cite{overweg_electrostatically_2018} and quantum dots \cite{eich_coupled_2018, banszerus_gate-defined_2018}. However, even in these devices, the natural flake edge may still provide conducting paths, giving rise to a large measured conduction that is independent of the nanostructured device. Furthermore, the effect of the electrostatic edges on the transport properties has remained largely unexplored.
\begin{figure*}[t]
    \includegraphics{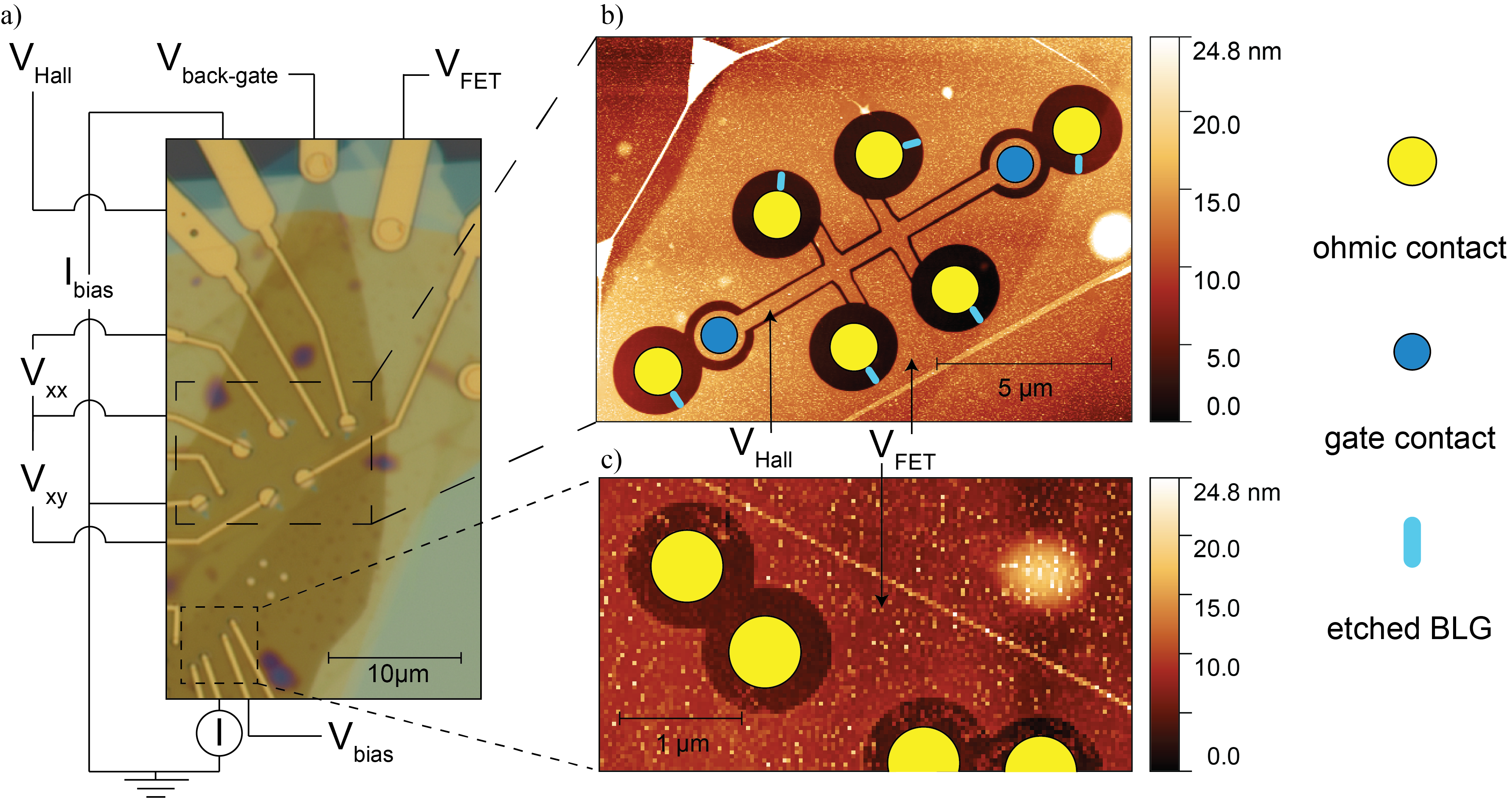}
    \caption{Measured devices. a) Optical microscope image of the finished devices and schematics of the setup. b) Atomic force micrograph of the Hall-bar after the patterning of the graphite top gate, in yellow the position of the ohmic contacts, in blue the contacts to the Hall-bar gate, and in light blue the cut in the BLG to facilitate equilibration. The Hall-bar gate is separated from the FET gate by a \SI{90}{\nano\meter}-wide gap. c) Atomic force micrograph of the \SI{800}{\nano\meter}-wide FET, in yellow the position of the ohmic contacts.}
    \label{fig:devices}
\end{figure*}

We have conceived a new architecture of dual-gated \gls{blg} devices facilitating the design of a large variety of them. The devices are fully surrounded by electrostatic edges, preventing any transport connection between ohmic contacts and natural flake edges, thus forcing the current to pass the device under study. Further, the active device area remains unaffected by exposure to polymers or high-voltage electron beams used in typical lithography processes. Taking into consideration the concerns about Cr and Ti sticking layers \cite{zheng_high-yield_2025} and following recent trends in the field \cite{zibrov_tunable_2017, noauthor_ultrasteep_nodate}, we move away from metal as top-gate electrode, but instead rely on graphite layers. 

After introducing the design architecture, we first demonstrate the cleanliness of the achieved band-gaps and the high mobility of the \gls{blg} charge carriers by performing magnetotransport measurements on a simple \gls{fet} device geometry. We then introduce a gate-defined Hall-bar, with which we characterize in more detail the remaining effects of the electrostatic edges in gate-defined devices.

\section{Device architecture}
Our device design relies on the band-gap that opens in \gls{blg} upon application of a perpendicular displacement field. In contrast to the successful demonstrations of gate-defined quantum point contacts and quantum dots in \gls{blg} \cite{overweg_electrostatically_2018,banszerus_gate-defined_2018, eich_coupled_2018, iwakiri_gate-defined_2022}, we don't rely on the band gap only in close vicinity of the active area. Instead, we tune the whole area of the \gls{blg} flake outside our active device into the resistive regime provided by the band-gap. Our device architecture is shown in Fig.~\ref{fig:devices}(b,c) in the form of a \gls{fet} and a Hall-bar device. The \gls{blg} flake is encapsulated between two \gls{hbn} flakes and, in an additional step, a few-layer graphite flake is stacked onto the top \gls{hbn} flake. The device geometry is then carved out of this graphite flake, separating the stack into the active device region and a region outside, which will be tuned into the band-gap using electrostatic gating. The latter effectively creates an electrostatic mesa. One more \gls{hbn} flake finally covers the graphite flake, allowing to contact all the graphite regions individually via small etched holes and a metal fan-out. Details on the stacking and device fabrication are described in section~\ref{sec:methods} and S1. The whole stack is visible in Fig.~\ref{fig:devices}(a).

The described architecture brings the advantage that conduction only occurs within the electrostatically defined mesa and, in particular, does not rely on the electrostatic opening of a clean band-gap along the flake edges. In contrast to previous \gls{blg} quantum dot designs \cite{eich_coupled_2018, banszerus_gate-defined_2018}, this prevents potential parallel conduction paths created by imperfect edges. For example, due to the way such edges are created by exfoliation, they are not necessarily uniform, and may partially consist of monolayer graphene, where no band-gap opens with a displacement field. Here, no conduction path connects our ohmic contact to the natural flake edges, thus overcoming this problem and increasing the device yield.

\section{Field-effect transistor}
\begin{figure*}[t]
     \includegraphics{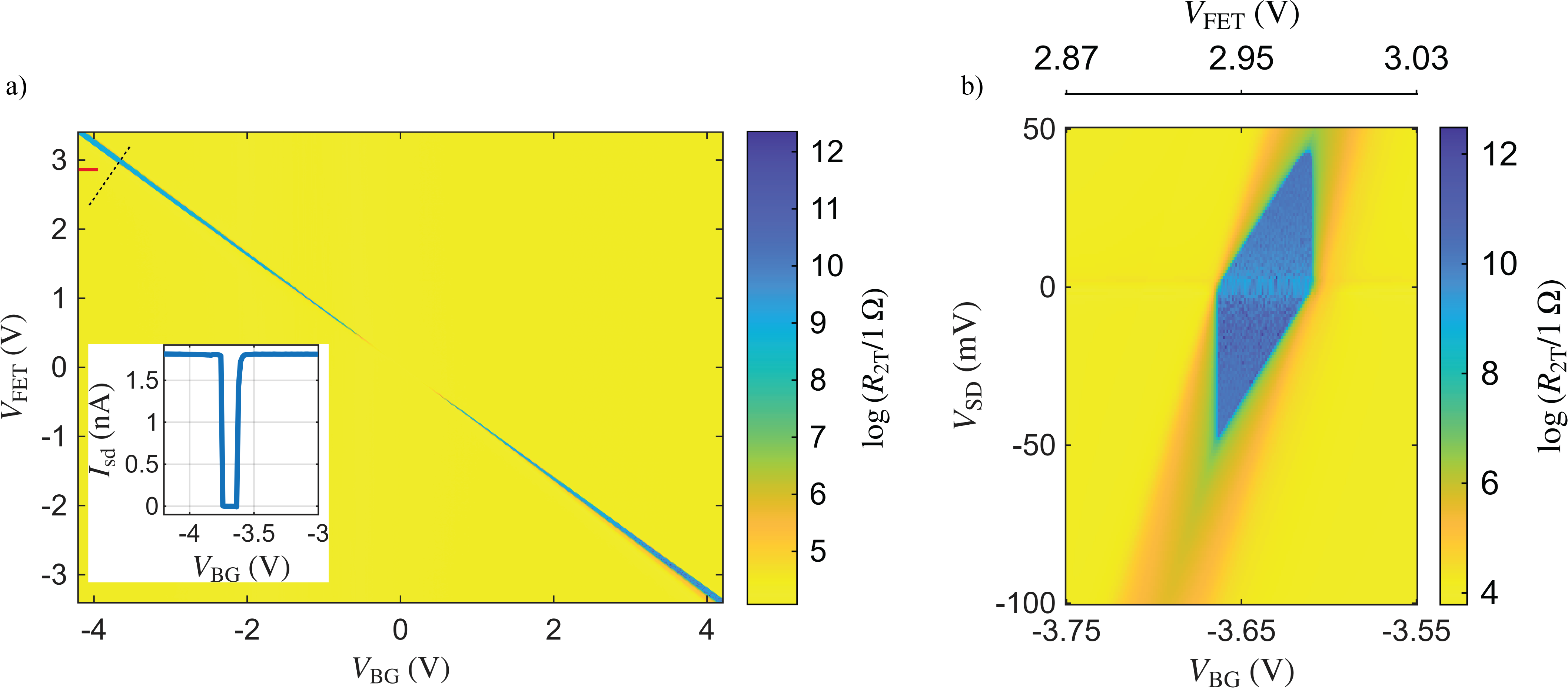}
    \caption{Band-gap measurements of the FET. a) Two-terminal resistance as a function of $\subtxt{V}{FET}$ and $\subtxt{V}{BG}$. The diagonal, highly insulating line appears when the Fermi level lies in the gap opened by the displacement field. Inset: measured current as a function of \subtxt{V}{BG} at a fixed top gate voltage as indicated by the red line.  b) Finite bias spectroscopy at fixed displacement field as indicated by the dotted line in a).}
    \label{fig:fet}
\end{figure*}
We first characterize our new architecture using a simple device design, namely a \gls{fet} as shown in Fig.~\ref{fig:devices}(c), which still allows us to study the magnetotransport properties of the \gls{blg} \cite{varlet_anomalous_2014, varlet_tunable_2015}. In our case, the \gls{fet} consists of two ohmic contacts separated by a graphite region which covers the whole surrounding. This excludes potential current paths along natural flake edges whenever the gated region is tuned into the band-gap, yielding the configurations of large resistance shown in Fig.~\ref{fig:fet}(a). There, the back-gate voltage \subtxt{V}{BG} and the graphite top-gate voltage \subtxt{V}{FET} are swept while a constant voltage $\subtxt{V}{SD} = \SI{750}{\mu V}$ is applied between the source and drain contacts. The resistance is large at the back-gate voltage \subtxt{V}{BG,0} and the top-gate voltage \subtxt{V}{TG,0} where the \gls{blg} is charge-neutral. The resistance reaches the \si{\tera\ohm} regime along the diagonal line, where the Fermi level lies in the band-gap opened by the perpendicular displacement field $D = \subtxt{C}{t} / 2\epsilon_0 (\subtxt{V}{TG} - \subtxt{V}{TG,0}) - \subtxt{C}{b} / 2\epsilon_0 (\subtxt{V}{BG} - \subtxt{V}{BG,0})$. Here, $C_i=\epsilon_0  \epsilon_i / d_i$, with $d_i$ and $\epsilon_i$ the \gls{hbn}  thickness and relative dielectric constant of the top ($i=\textrm
t$) or bottom ($i=\textrm{b}$) flake and $\epsilon_0$ the vacuum dielectric constant, respectively.

To test the uniformity and homogeneity of the band gap, we extract the resistance at constant displacement field, while tuning the Fermi energy through the band-gap by changing the density $n = \subtxt{C}{t}/ e\left(\subtxt{V}{TG} - \subtxt{V}{TG,0}\right) + \subtxt{C}{b} / e\left(\subtxt{V}{BG} - \subtxt{V}{BG,0}\right)$ from positive (hole charge carriers) to negative (electron charge carriers). Meanwhile, we step \subtxt{V}{SD} in order to measure the size of the band-gap \cite{icking_transport_2022}, as shown in Fig.~\ref{fig:fet}(b). The uniform diamond with its sharp transition from a high in-gap resistance to low-resistive conduction outside the gap suggests a complete suppression of transport channels in the insulating regime. These results demonstrate the absence of common fabrication-related issues named above, which typically lead to impurities and defects in the measured region of the flake.

We now characterize the magnetotransport of the \gls{fet} in the conducting regime, when the Fermi level lies in the valence or conduction band. In the following measurements, we either work at constant displacement field or constant density around operation points \subtxt{V}{BG}$^*$ and \subtxt{V}{TG}$^*$. For this, the top-gate and back-gate have to be moved simultaneously following the relations 
\begin{align}
\subtxt{V}{TG} - \subtxt{V}{TG}^* &= -\frac{C_b}{C_t}\,\left(\subtxt{V}{BG} - \subtxt{V}{BG}^*\right)
\quad \text{(for } n=\text{const)} \\
\subtxt{V}{TG} - \subtxt{V}{TG}^* &= \phantom{-}\frac{C_b}{C_t}\,\left(\subtxt{V}{BG} - \subtxt{V}{BG}^*\right)
\quad \text{(for } D=\text{const)}
\end{align},
with the capacitance ratio extracted from the measurements in Fig.~\ref{fig:fet}(a). We again apply a constant voltage $\subtxt{V}{SD} =  \SI{125}{\micro \volt}$ and measure the current flowing between the contact pair in a two-terminal setup. With the chosen \gls{fet} design, as soon as the states in the gated regime are quantized into Landau levels, edge channels form and surround the holes in the graphite layer defining the contact area. Thus, at constant filling factor, the device becomes insulating. However, current flows when the Fermi energy in the gated region is close to a Landau level, within its broadening. The measured resistance is equivalent to the longitudinal resistance expected in a Corbino geometry. The etched regions without top-gate, which directly surround the contacts, remain conductive at all magnetic field values, as etch inhomogeneities S2 prevent Landau quantization. This effect is important in our measurements, as it allows us to probe the Landau levels in the gated region.
\begin{figure*}[t]
    \includegraphics{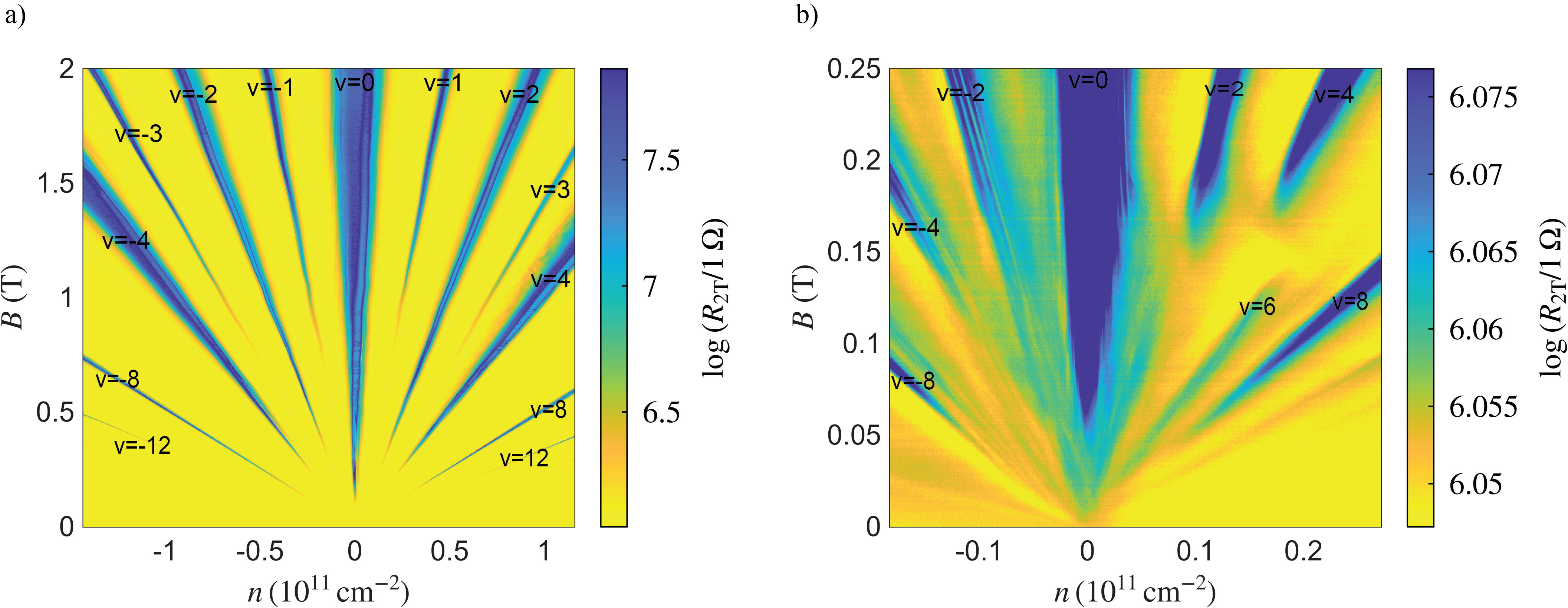}
    \caption{Magnetotransport measurement of the FET at $D=\SI{0}{\volt/\nano \meter}$. a) Landau fan from $\SI{0}{\tesla}$ to $\SI{2}{\tesla}$, with the visible filling factors labeled in black. b) Zoom-in of the low field and low density region.}
    \label{fig:fetLandau}
\end{figure*}
We set the displacement field to $D = 0$ and measure a Landau fan as shown in Fig.~\ref{fig:fetLandau}. The breaking of the fourfold symmetry due to spin and valley is clearly seen up to filling factor $\nu = 12$ at magnetic fields $B\leq\SI{0.5}{\tesla}$. This indicates long quantum lifetimes $\subtxt{\tau}{q}$ causing the broadening $\subtxt{\Gamma}{LL} \sim \hbar / \subtxt{\tau}{q}$ of Landau levels to be much lower than their separation even at these relatively low fields. The separation of orbital levels is determined by the cyclotron frequency to $\hbar \subtxt{\omega}{c}$, while the Zeeman energy $g_i\mu_B B$ sets the separation of spin- or valley-split levels, with $g_i$ the spin- or valley g-factor, respectively. The spin-splitting between the first and second Landau level is visible down to $B = \SI{200}{\milli\tesla}$, and we resolve \gls{sdh} oscillations down to $B = \SI{4}{\milli\tesla}$ for filling factor $\nu = -8$, already identified as the most robust filling factor at low magnetic fields in previous studies \cite{seiler_probing_2024}. Following these observations, we estimate a quantum lifetime of $\subtxt{\tau}{q} \sim \SI{47}{\pico\second}$ and a quantum mobility of $\subtxt{\mu}{q} \sim \SI{2500000}{\centi\meter^2/\volt\second}$.

\section{Hall-bar}
We make use of the flexibility in device design provided by our new architecture to produce a fully gate-defined Hall-bar, visible in Fig.~\ref{fig:devices}(b). The displacement field and the density in the Hall-bar are tuned by the voltage \subtxt{V}{Hall} applied to the central graphite gate. The outside region of the Hall-bar is the same graphite region as the \gls{fet} gate, hence it is tuned by \subtxt{V}{FET}. Using the results of the \gls{fet} measurements, we keep the Fermi energy in that outside region in the band gap. In this setting, we sweep \subtxt{V}{BG} versus \subtxt{V}{Hall} in order to determine the capacitance of the Hall-bar gate on the \gls{blg} underneath. We measure resistances in the \si{\giga\ohm} regime, again indicating a high device quality with a clean band-gap,  and also implying that no transport occurs underneath the ungated \SI{90}{\nano\meter} gap between the Hall-bar gate and the \gls{fet}-gate.

\begin{figure*}[t]
    \includegraphics{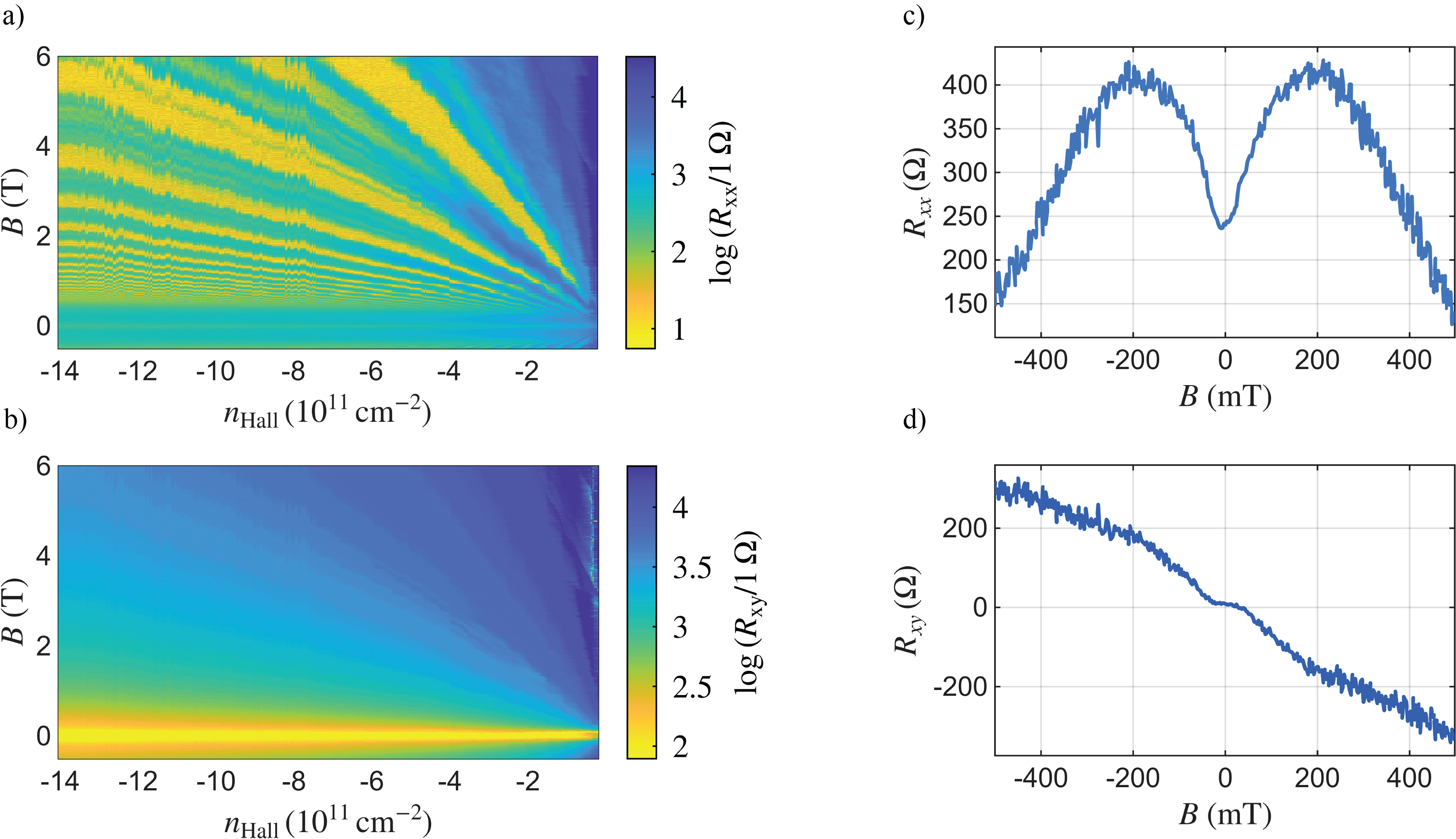}
    \caption{Magnetotransport measuremenet of the Hall-bar at $D=\SI{0.1}{\volt~/\ \nano \meter}$. a) Landau fan in $\subtxt{R}{xx}$. b)  Landau fan in $\subtxt{R}{xy}$. c) Low-field trace of $\subtxt{R}{xx}$ taken at $n_{Hall}=\SI{1.23e12}{\centi\meter^{-2}}$, showing the mesoscopic magnetoresistance peaks. d) Low-field trace of $\subtxt{R}{xy}$ taken at $n_{Hall}= \SI{1.23e12}{\centi\meter^{-2}}$, showing the quenching of the Hall resistance.}
    \label{fig:hbLfan}
\end{figure*}

We apply a magnetic field perpendicular to the \gls{blg} flake to measure the transverse and longitudinal resistances $\subtxt{R}{xy}$ and $\subtxt{R}{xx}$ shown in Fig.~\ref{fig:hbLfan}(a) and (b), respectively. The typical signatures of the integer quantum Hall effect and \gls{sdh} oscillations are visible at magnetic fields above about \SI{0.5}{\tesla}. The low-field data shown in Fig~\ref{fig:hbLfan}(c) and (d) displays an increase of $\subtxt{\rho}{xx}$ with magnetic field up to a peak, followed by a decreases, while $\mathrm{R_{xy}}$ exhibits a decreased slope around zero field. The flattening of the transverse resistance occurs due to the quenching of the Hall effect in the ballistic regime, when the Lorentz force does not bend the trajectories of the injected carriers enough for them to be collected by the neighboring voltage probes \cite{roukes_quenching_1987, thornton_ballistic_1998}.

\begin{figure*}[t]
    \includegraphics{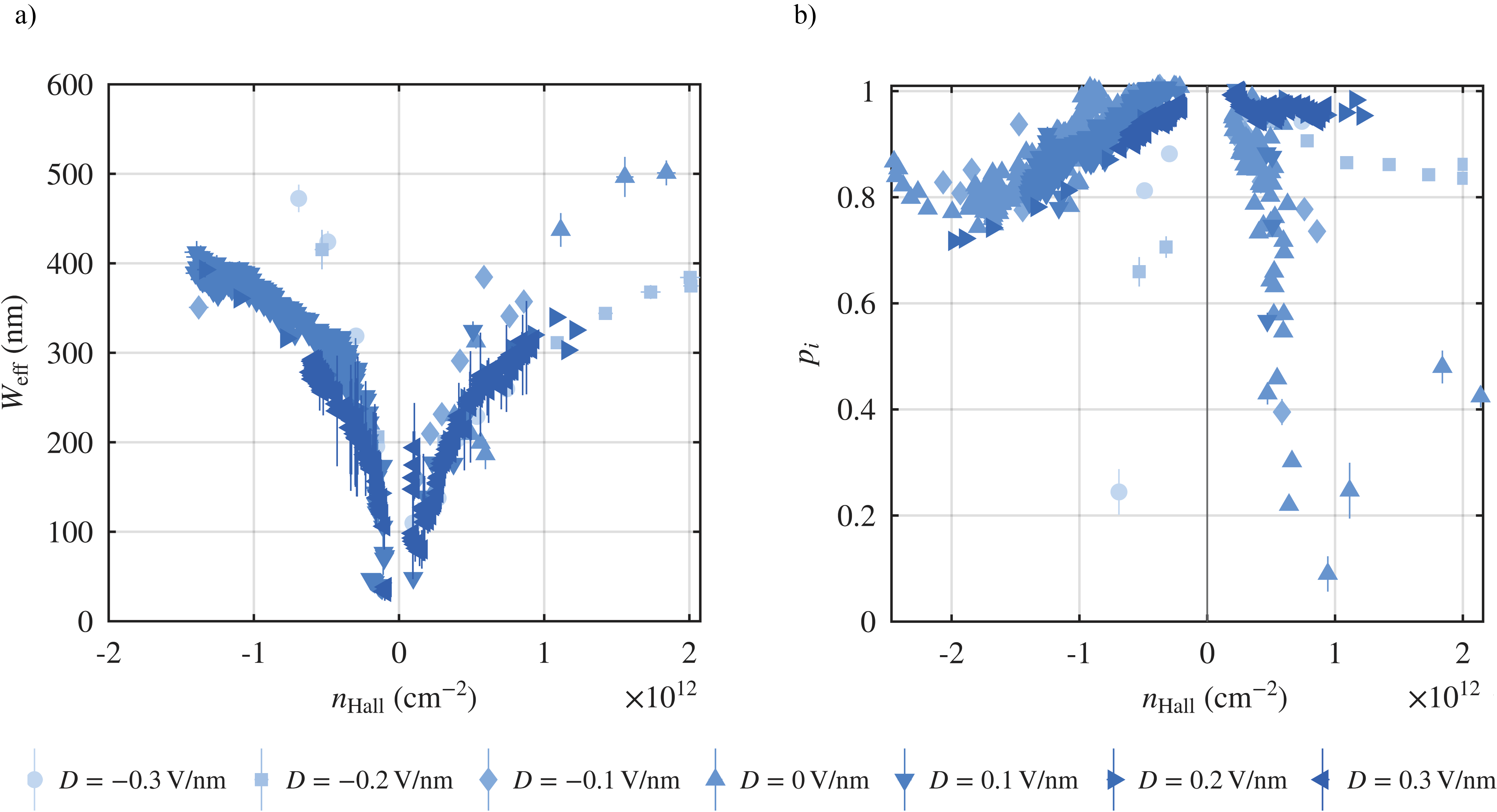}
    \caption{Hall-bar analysis. a) Effective width $\subtxt{W}{eff}(n)$ extracted as a function of density and plotted for different displacement fields. b) Scattering parameter $p_i(n)$ calculated for each density and plotted for different displacement fields.}
    \label{fig:hbWeff}
\end{figure*}

We associate the increase in the longitudinal resistance to the narrow channel compared to the mean free path as estimated further below. In this mesoscopic regime, specular edge scattering in combination with cyclotronic motion deflect the charge carriers back to the source contact \cite{soffer_statistical_1967, thornton_ballistic_1998}. Only at larger fields, when the cyclotron orbit becomes smaller than the width of the Hall-bar, the carriers are forward scattered and the magnetoresistance decreases again.
The thin metal theory \cite{soffer_statistical_1967, fuchs_conductivity_1938-1, sondheimer_mean_2001} yields the value of the magnetic field \subtxt{B}{max} at which the longitudinal resistance peaks as
\begin{equation}
    \subtxt{B}{max}=0.55\frac{\hbar k_\mathrm{F}}{eW}
      \label{Eq:B_peak}
\end{equation}
where $\hbar$ is the reduced Planck constant, $e$ is the electron charge, $W$ is the channel width, and $k_\mathrm{F}=\sqrt{\frac{4}{g}\pi n}$ is the Fermi wave vector. In \gls{blg}, the fourfold spin- and valley degeneracy leads to $k_\mathrm{F}=\sqrt{\pi n}$. From this equation, we determine the effective width $\subtxt{W}{eff}$ of our Hall-bar channel as a function of density and displacement field, as shown in Fig.~\ref{fig:hbWeff}(a). The data follows a power-law relation, $W_\mathrm{eff}\propto {n}^\beta$, consistent with a near-parabolic confinement potential, with $\beta$ being weakly dependent on the displacement field (see Supplementary Information).
The low-field behavior of the longitudinal resistivity $\subtxt{\rho}{xx}$, which we now extract from \subtxt{R}{xx} using the effective width, is dominated by edge scattering. In particular, the mean of $\subtxt{\rho}{xx}$ measured in our devices at low fields deviates from its zero-field value, a behavior that differs from the characteristic response observed in large devices. This allows us to further analyze the edge scattering. We extract $\subtxt{\rho}{0, avg}$ from the mean of the signal measured at fields above \SI{0.5}{\tesla} but below Landau quantization and compare it with the measured value $\subtxt{\rho}{0,meas}$ measured at zero field. The measured deviation (see Supplementary Information, S6) can be parametrized by a scattering parameter $p$ \cite{soffer_statistical_1967, thornton_ballistic_1998} via
\begin{align}
    \rho_\mathrm{eff} & =\rho_0+\rho_0 l_0 \frac{(1-p)}{W} = \rho_0+\frac{\hbar k_\mathrm{F}}{ne^2} \frac{(1-p)}{W} \\
    \Leftrightarrow p&=1-\frac{{e^2}}{\hbar\sqrt{\pi} }\frac{(\subtxt{\rho}{0,meas} - \subtxt{\rho}{0,avg})W_\mathrm{eff}^2\sqrt{n}}{L}
\end{align}
Here, $\rho_\mathrm{eff}$ is the effective resistivity of the channel, $\rho_0 l_0=\hbar k_\mathrm{F} / ne^2$, $L$ is the distance between the Hall voltage probes, and in the last line, we replaced the sample width $W$ with the density-dependent width \subtxt{W}{eff}. The scattering parameter takes values between $0$, indicating completely diffusive scattering, and $1$ denoting exclusive specular scattering.
The data in Fig.~\ref{fig:hbWeff}(b) shows a trend from predominant specular scattering at low densities to slightly increased diffusive scattering at high density. To gain some insight, let us first comment on the mesoscopic regime we are working in. Even though the Hall-bar width is of the same order of magnitude as the Fermi wavelength, we do not observe quantization effects typical for a quantum point contacts. We suspect that only a few modes are occupied, which are mixed by scattering and thus cannot be resolved individually. 
\begin{figure}[h]
\centering
    \includegraphics[width=0.7\columnwidth]{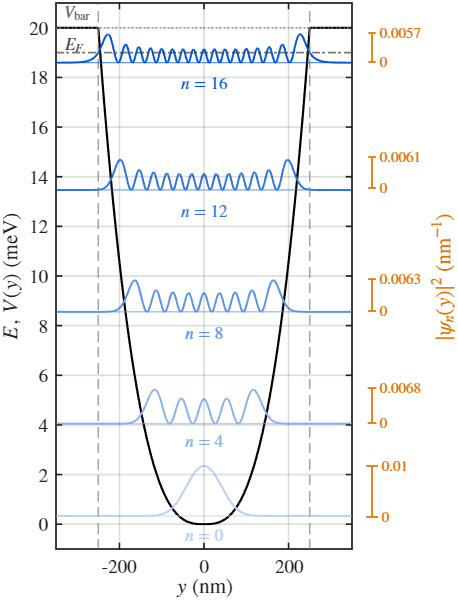}
    \caption{Transverse confinement potential and quantized modes. The vertical dashed lines indicate the geometric barrier edges 
    while the horizontal dotted line marks the barrier height $V_{\mathrm{bar}}$. The calculated transverse eigenenergies $E_n$ are shown as horizontal levels for each mode number $n = 0, 4, 8, 12, 16$. The corresponding probability density $|\psi_n(y)|^2$, scaled for visualization, is plotted with a vertical offset set by its eigenenergy $E_n$, with a local probability scale shown alongside each mode (orange). The dash-dotted horizontal line indicates the Fermi energy $E_F$. Modes approaching the barrier energy exhibit an increasingly extended transverse probability distribution and enhanced penetration into the classically forbidden region.}
    \label{fig:conf}
\end{figure}
Nevertheless, we rationalize the observed evolution of $p$ by means of the wave-functions expected in a one-dimensional near-parabolic confinement potential, as shown in Fig.~\ref{fig:conf} .
Using this model, we note that the probability amplitude of the wavefunctions at the lithographic edge of the Hall-bar is close to zero for the ground and first excited states, consistent with the extracted $p$ close to $1$ at low densities. With increasing mode number, the wavefunction amplitudes start to be finite at the lithographic Hall-bar edge, in agreement with $p$ starting to decrease at higher density. This suggests that the mixing of the subbands is caused predominantly at the edges, while the bulk supports long mean free paths in accordance with the \gls{fet} measurements.

Using the insights obtained so far, we now estimate the carrier mobility using the resistivity $\subtxt{\rho}{0,avg}$. The Hall mobility extracted at densities above \SI{2e11}{cm^{-2}} amounts to \SI{200000}{cm^2/Vs}, comparable with state-of-the-art devices \cite{icking_transport_2022, iwakiri_gate-defined_2022, kumar_quarter-_2025}. The corresponding mean free path of about \SIrange{1}{3}{\micro\meter} exceeds the Hall-bar length, suggesting that the diffusive model used to extract transport mobilities is not suitable for describing our device.  
Nevertheless, the temperature-dependent \gls{sdh} oscillations (see Supplementary Information, S7) allow us to extract a value for the quantum mobility of roughly \SI{30000}{{\centi\meter^2/\volt\second}}, much lower than the one found for the \gls{fet}, indicating that the edges in the Hall-bar induce small-angle scattering.

\section{Origin of the disorder}
\label{sec:origindisorder}
The \gls{fet} measurements revealed record high values of the quantum mobility, indicating exceptional cleanliness of our flakes. Compared to that, the Hall-bar showed significant edge scattering, demonstrating the importance of understanding the electrostatic edges of gated device designs. In our device, the edges are defined by the small gap etched into the graphite top gate, separating the Hall-bar gate from the FET-gate. We, therefore, set out to analyze in more detail the effect of the applied etching process.
\begin{figure*}[t]
    \includegraphics{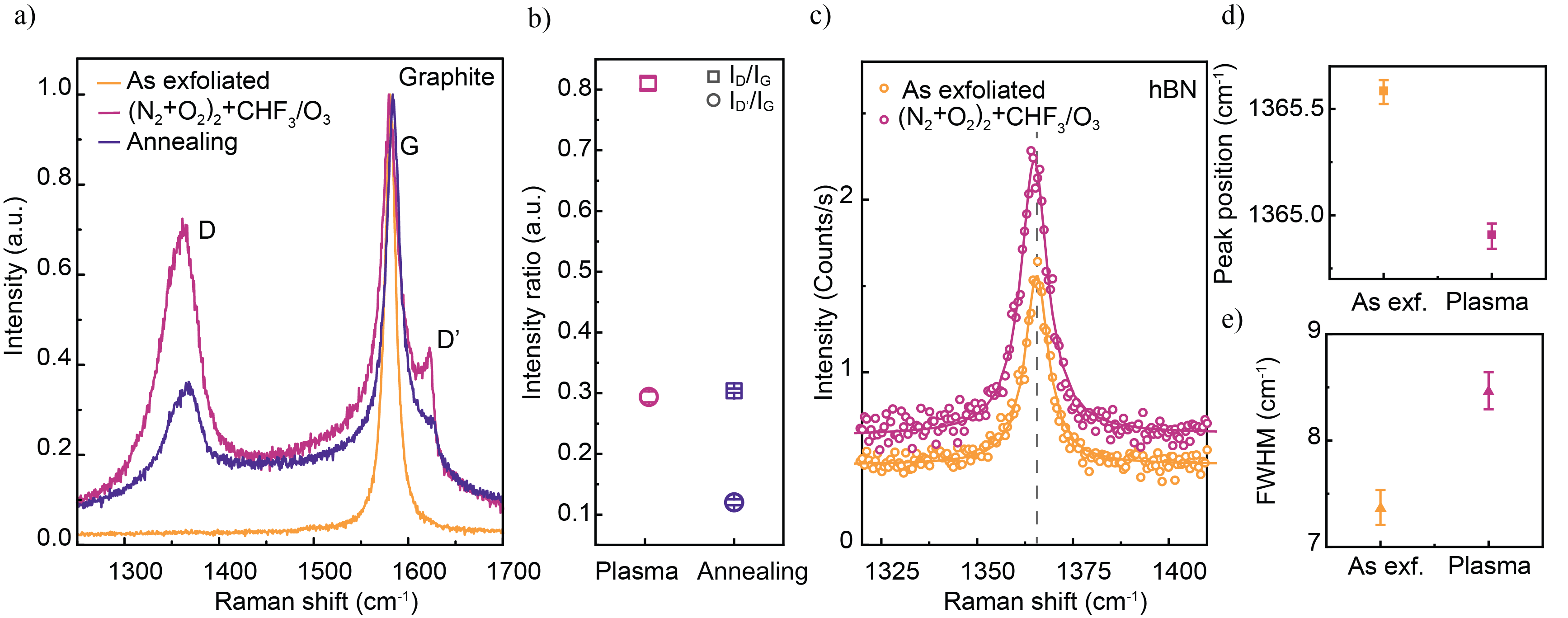}
    \caption{Effect of the plasma treatment on a thin graphite and \gls{hbn} flake, respectively. a) Raman spectra of as-exfoliated (orange), plasma-treated (pink) and annealed (purple) graphite flake on top of $\mathrm{SiO_2/Si}$. b) Intensity of the D and D' peaks relative to the G-peak, $\mathrm{I_D/I_G}$ and $\mathrm{I_{D'}/I_G}$, respectively, used to quantify the disorder after plasma treatment and annealing. c) Raman spectra of as-exfoliated (orange) and plasma-treated (pink) \gls{hbn} flake on top of $\mathrm{SiO_2/Si}$. d) Peak position and e) full width at half maximum of the $\mathrm{E_{2g}}$ mode before and after plasma treatment.}
    \label{fig:RamanGraphene}
\end{figure*}
We test the transport properties of the \gls{blg} in regions where the top graphite gate has been etched (see Supplementary Information). Magnetotransport measurements show that resistance peaks corresponding to Landau level gaps are minimally pronounced, indicating the presence of disorder in these regions.

Further, we perform Raman and \gls{kpfm} measurements to analyze the crystal quality and the local surface potential of \gls{hbn} and graphite flakes treated with plasma etching. For this, \gls{hbn} and graphite flakes are exfoliated on $\mathrm{SiO_2}$ substrates and treated with a plasma process essentially equal to the one used for our device fabrication. It consists of two (instead of four used in the device) cycles of \SI{20}{\second} of$\mathrm{N_2}$  plasma at \SI{100}{\watt} and \SI{40}{\second} of $\mathrm{O_2}$ plasma at \SI{60}{\watt} with a \SI{10}{\second} $\mathrm{CHF_3/O_2}$ plasma step at \SI{60}{\watt}. 


The Raman measurements performed on the graphite sample are shown in Fig.~\ref{fig:RamanGraphene}(a). The orange solid line represents the Raman spectrum of the graphite flake before being exposed to the plasma treatment. The measured intensity exhibits only one narrow peak at around \SI{1582}{\centi\meter{^{-1}}} attributed to the G-band that arises from in-plane motion of the carbon atoms in bulk graphite. All the spectra are normalized to the intensity measured at the respective G-peak. After the plasma treatment,
two more peaks appear in the Raman spectrum, as seen in the pink solid line in Fig.~\ref{fig:RamanGraphene}(a): the D-band and D'-band, around \SI{1350}{\centi\meter{^{-1}}} and \SI{1620}{\centi\meter{^{-1}}} respectively. Both peaks are attributed to defects in the graphite \cite{dias_production_2016}, and their intensity relative to the G-band is commonly used to evaluate the defect density in graphitic materials \cite{madito_revisiting_2025, eckmann_probing_2012}. In our case, we attribute their appearance to defects, edge disorder, and contaminants introduced by the plasma treatment. 

We employ an annealing treatment, \SI{5}{\hour} \SI{350}{\celsius} in a $\mathrm{N_2}$/$\mathrm{H_2}$ atmosphere under a pressure of \SI{60}{\milli\torr}, to 
to test whether the observed changes are reversible.
The solid purple line in Fig.~\ref{fig:RamanGraphene}(a) shows the Raman spectrum acquired after the annealing procedure. 
To facilitate the comparison of the D and D' peaks before and after annealing, we plot their intensity relative to the G-peak in Fig.~\ref{fig:RamanGraphene}(b). Clearly, the relative intensity of both peaks decreases after annealing, indicating a partial removal of the plasma-induced disorder localized at the sample surface. 
The reduction in disorder is also inferred from the evolution of the G-band linewidth, which increases after the plasma exposure and subsequently decreases after the annealing treatment, as qualitatively seen in Fig.~\ref{fig:RamanGraphene}(a) and quantitatively analyzed in Fig. S15(b) of the Supplementary Information. However, even after annealing, the D and D' peaks persist and the G-band's linewidth remains somewhat larger, suggesting that the plasma treatment also introduces permanent structural defects that remain after the used annealing treatment. 
In addition, the ratio of the D versus D' peak gives insights into the defect type \cite{eckmann_probing_2012}. The measured ratio of around \num{3} before and after annealing points to boundary-like defects, such as, for example, grain boundaries \cite{madito_revisiting_2025}.

Also the \gls{hbn} flake was first measured as exfoliated and then after the plasma treatment, with Raman spectra shown in Fig.~\ref{fig:RamanGraphene}(c).
The observed peak around \SI{1366}{\centi\meter^{-1}} is characteristic for \gls{hbn} and corresponds to the $\mathrm{E_{2g}}$ phonon mode \cite{nemanich_light_1981, De_Luca_2020}. After the plasma treatment, the $\mathrm{E_{2g}}$ peak exhibits a red shift, as shown in Fig.~\ref{fig:RamanGraphene}(d), which can be attributed to a modification of the \gls{hbn} lattice \cite{radhakrishnan_fluorinated_2017, na_modulation_2021}. In our case, the plasma treatment might introduce strain, defects, and a reduction of the \gls{hbn} thickness. The reduction of thickness is further supported by the change in the optical contrast of the flake, as shown in  Fig. S16(b) of the Supplementary Information. 
In addition to the red shift, the $\mathrm{E_{2g}}$ peak becomes wider, as shown in Fig.~\ref{fig:RamanGraphene}(e). This indicates a reduction of the phonon lifetime due to additional scattering channels arising, for example, from the defects. Also here, we attempt at reversing plasma-induced defects with an annealing treatment. However, the observed signatures point towards increased contamination, possibly due to redistributing residues from the exfoliation tape (see Supplementary Information).


To get insights into whether the induced disorder arises from the oxygen and fluorinated plasmas \cite{sevak_singh_band_2014, na_modulation_2021,radhakrishnan_fluorinated_2017, meiyazhagan_gas-phase_2021, cho_improvement_2014}, we perform similar measurements using a different etching process. Namely, we replace the final $\mathrm{CHF_3/O_2}$ step with a \SI{10}{\second} $\mathrm{N_2}$ plasma treatment at \SI{100}{\watt}, which should leave a clean and inert surface of the \gls{hbn} flake \cite{ma_control_2019}. The Raman measurements did not show significant differences, as presented in Fig. S19 of the Supplementary Information.

In addition to the Raman measurements, \gls{kpfm} measurements provide insight into the work function differences of the substrates under study. We compare two samples with \gls{hbn} flakes exfoliated on $\mathrm{SiO_2}$ chips, one of them exposed to the $\mathrm{CHF_3/O_2}$ plasma process, and one to the $\mathrm{N_2}$ process. In the former, we observe a broader and stronger surface-potential perturbation around the flakes, indicating larger work function differences between the exposed \gls{hbn} and $\mathrm{SiO_2}$. Comparably, the second recipe results in a more homogeneous potential landscape. These observations are in agreement with literature \cite{radhakrishnan_fluorinated_2017, meiyazhagan_gas-phase_2021, cho_improvement_2014}. The measurements do not allow to differentiate whether the two recipes attack the \gls{hbn} flake in distinct ways, or whether the observed differences are confined to the boundaries of the flake, which are known to be affected by plasma processes in general \cite{sevak_singh_band_2014, na_modulation_2021}.


In view of the above findings and considering that the patterned graphite is completely removed, as well as a few layers of the \gls{hbn}, there are two different but potentially coexisting explanations for the disorder observed in the transport measurements. Possibly, the \gls{hbn} gets disordered and structurally damaged by the etching process, inducing a disordered potential landscape in the \gls{blg}. Alternatively, the disorder in the potential landscape originates from the contaminated and potentially functionalized edges of the etched graphite gates.

\section{Conclusion}
We introduce a new architecture for graphite-gated \gls{blg} devices that enables the definition of active regions completely isolated from any natural flake edge. This architecture yields a highly resistive and clean band gap, together with a large quantum mobility. It also enables new device geometries, and we demonstrate the straightforward realization of a gate-defined Hall-bar, a geometry that has so far received limited attention in graphite-gate-defined \gls{blg} devices.

In contrast to a \gls{fet}, the Hall-bar geometry allows for the measurement of both the longitudinal and transverse resistances, yielding complementary information. In our device, the Hall-bar operates in the mesoscopic regime, with the mean free path comparable to the device dimensions. We are therefore sensitive to transport phenomena such as Hall-effect quenching and boundary scattering. This allows us to investigate the role of electrostatically defined edges, an aspect of gate-defined devices which has not been addressed so far.

We find that the mobility in the Hall-bar measurements is limited by scattering associated with disorder in the electrostatic potential. Raman spectroscopy and \gls{kpfm} measurements indicate that this originates from defects and disorder introduced by the fabrication process. Meanwhile, measurements of the \gls{fet} demonstrate the extraordinary quality of the bulk, with a quantum mobility approaching \SI{2.5e6}{\centi\meter^2/\volt\second}. 
\section{Methods}
\label{sec:methods}

\textbf{Devices fabrication}. The devices were fabricated using a stack comprising a graphite back gate, \gls{hbn}, bilayer graphene, six-layer graphite, and \gls{hbn}.
The stack was assembled with a standard dry transfer technique. 
The top-gate patterning involved standard lithographic techniques followed by an \gls{rie} step. The etching process consisted of four cycles of \SI{20}{\second} $\mathrm{N_2}$  plasma at \SI{100}{\watt} and \SI{40}{\second} $\mathrm{O_2}$ plasma at \SI{10}{\watt}, followed by $10\mathrm{s}$ of $\mathrm{CHF_3/O_2}$ plasma at $60\mathrm{W}$. 
Edge ohmic contacts and Hall-bar gate contacts were realized in the same fabrication step, in which a lithographic mask was used to selectively etch the \gls{hbn} by \gls{rie} with SF$_6$ plasma at $6\mathrm{W}$, followed by $15\mathrm{s}$ of $\mathrm{CHF_3/O_2}$  plasma at $60\mathrm{W}$ to etch the unaffected bilayer graphene and expose its edges. The \gls{rie} step was followed by metal evaporation of a Cr/Pd/Au stack with thicknesses of $10\mathrm{nm}$, $30\mathrm{nm}$, and $40\mathrm{nm}$, respectively.\\

\textbf{Measurements}.
All the measurements were performed in a BlueFors dilution cryostat at a base temperature of $10\mathrm{mK}$. 
Current-biased measurements were performed by applying a voltage and placing a $200\mathrm{M\Omega}$ resistor in series with the device. The current was measured using a current-to-voltage converter connected to a lock-in amplifier. Lock-in measurements were performed using three synchronized Zurich Instruments lock-in amplifiers to measure \subtxt{V}{xx}, \subtxt{V}{xy} and the current. The voltages $\subtxt{V}{xx}$ and $\subtxt{V}{xy}$ were amplified using differential voltage amplifiers before being fed into the lock-in amplifiers.\\

\textbf{Raman spectroscopy}. 
Raman spectra were acquired using a 515 nm COBOLT laser with a power of 1 mW for the graphite samples and 400 $\mathrm{\mu W}$ m for the \gls{hbn} samples. The laser was focused onto the samples using a 50x objective with a numerical aperture of 0.60. The Raman spectra were collected with a Horiba T6400 triple monochromator with a spectral resolution of 0.5 $\mathrm{cm^{-1}}$ and equipped with a $\mathrm{N_2}$ cooled charge coupled device (CCD) detector. The Raman spectra of the graphite samples were collected with an integration time of 60 s and averaged over 3 scans, whereas the Raman spectra of the \gls{hbn} samples were acquired with an integration time of 120 s and averaged over 3 scans.

\section*{Acknowledgments}
We thank Michael Steinacher for his valuable contributions to the optimization of the measurement setup. We thank Clara Galante, Jessica Richter, Tijl Degroote, Max Ruckriegel, Markus Niese, Marta Perego and Artem Denisov for helpful input and discussions around building first stacks and exploring bilayer graphene physics.

We acknowledge financial support from the Swiss State Secretariat for Education, Research and Innovation, the Quantum Transitional Call, and from the Swiss National Science Foundation through grant number 200021-231373.


\section*{Data availability}
The data that support the findings of this study are openly available in ZENODO at https://zenodo.org/doi/xxx.


\bibliography{arXiv_Submission_V1}
\bibliographystyle{apsrev4-2}

\begin{figure*}[p]
    \centering
    \includegraphics[page=1,width=\textwidth]{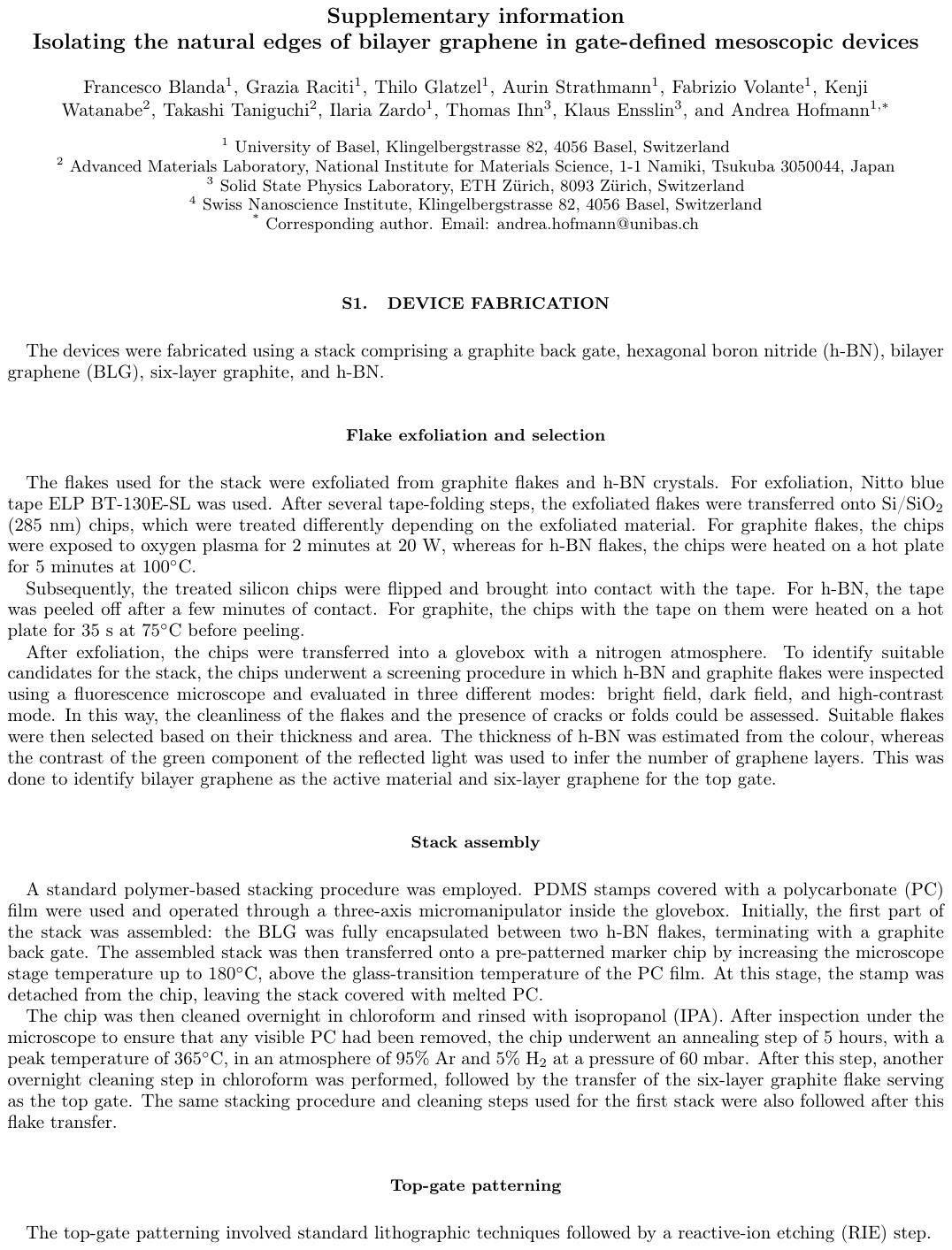}
\end{figure*}

\begin{figure*}[p]
    \centering
    \includegraphics[page=2,width=\textwidth]{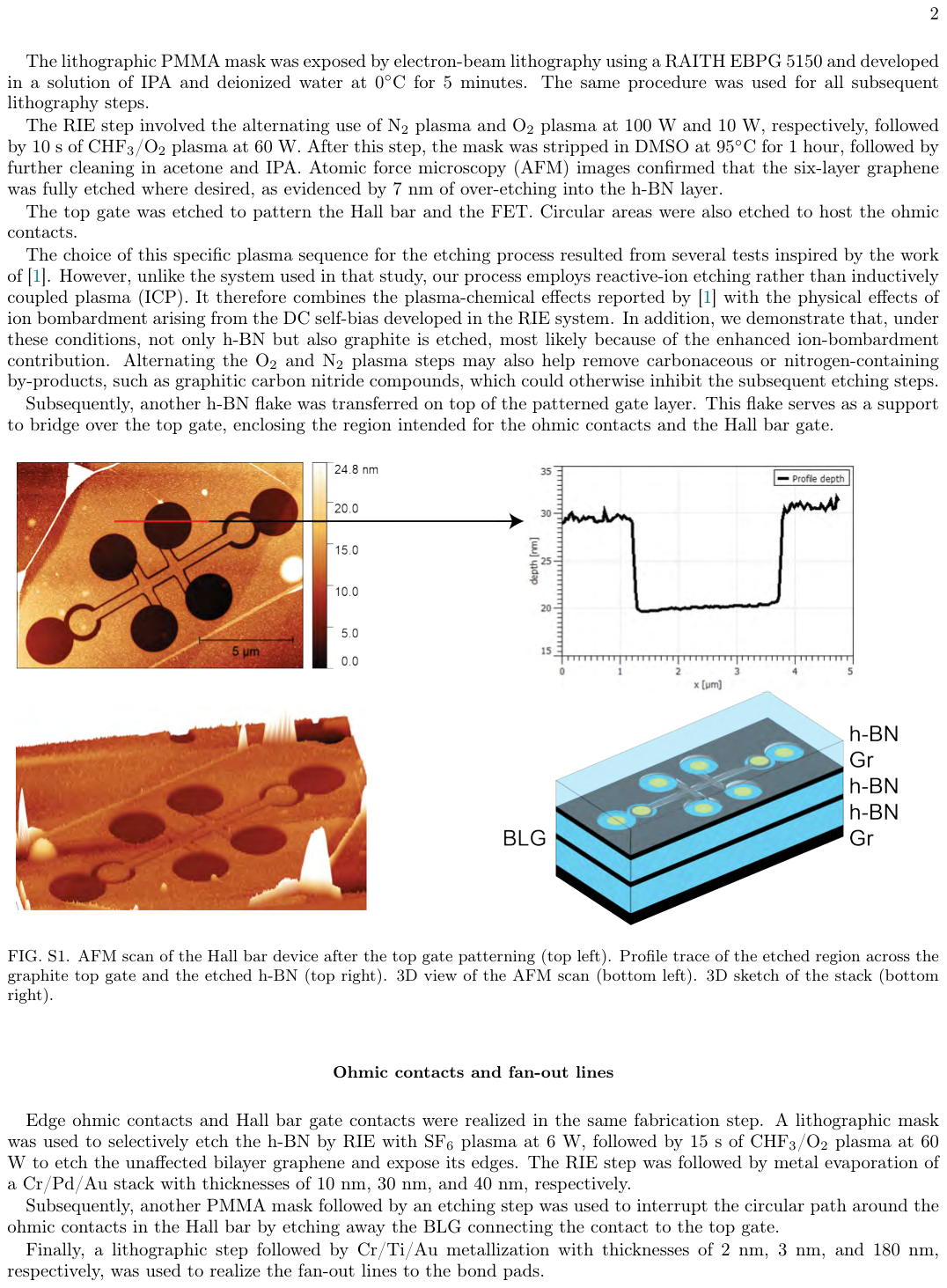}
\end{figure*}

\begin{figure*}[p]
    \centering
    \includegraphics[page=3,width=\textwidth]{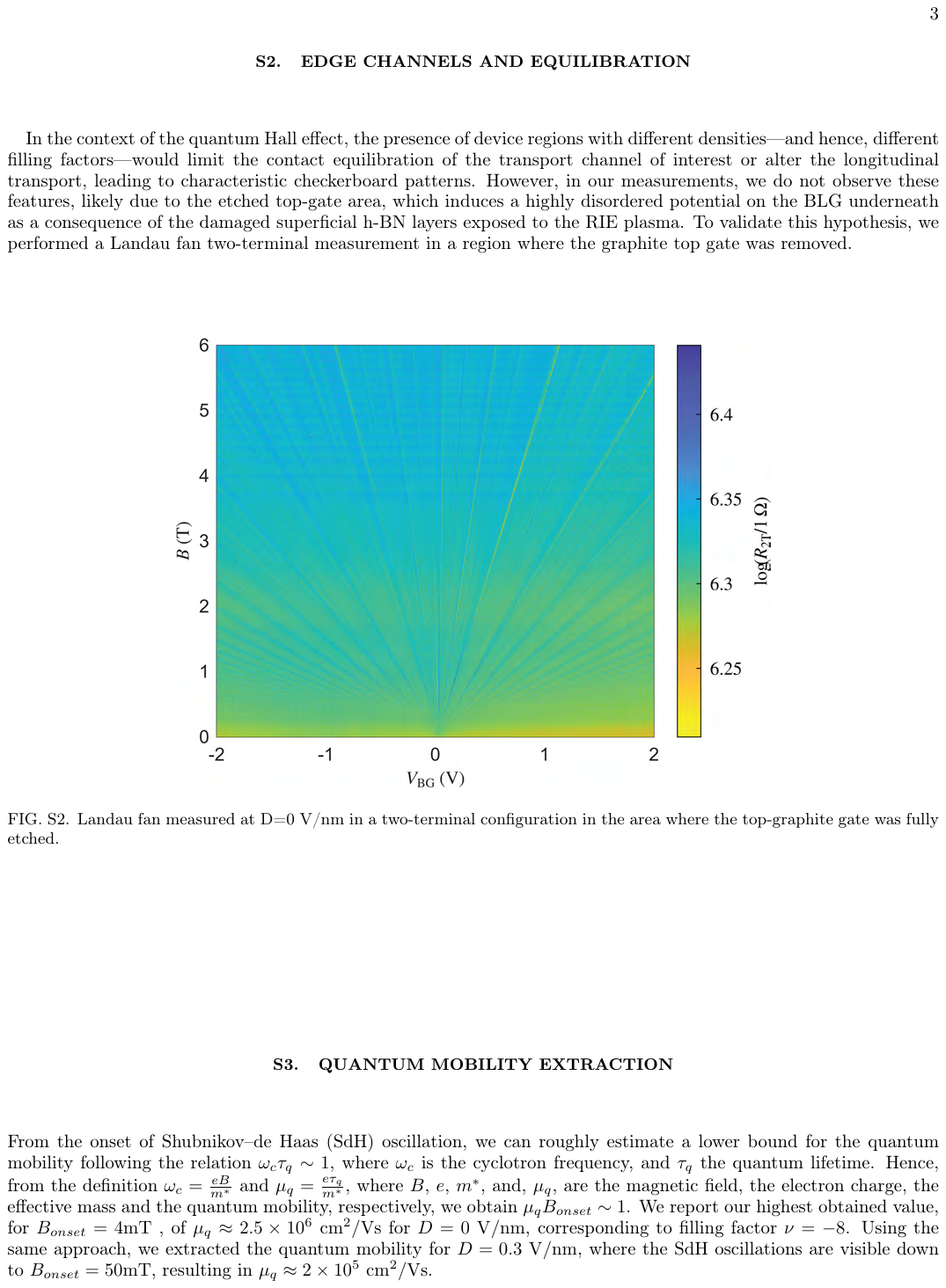}
\end{figure*}

\begin{figure*}[p]
    \centering
    \includegraphics[page=4,width=\textwidth]{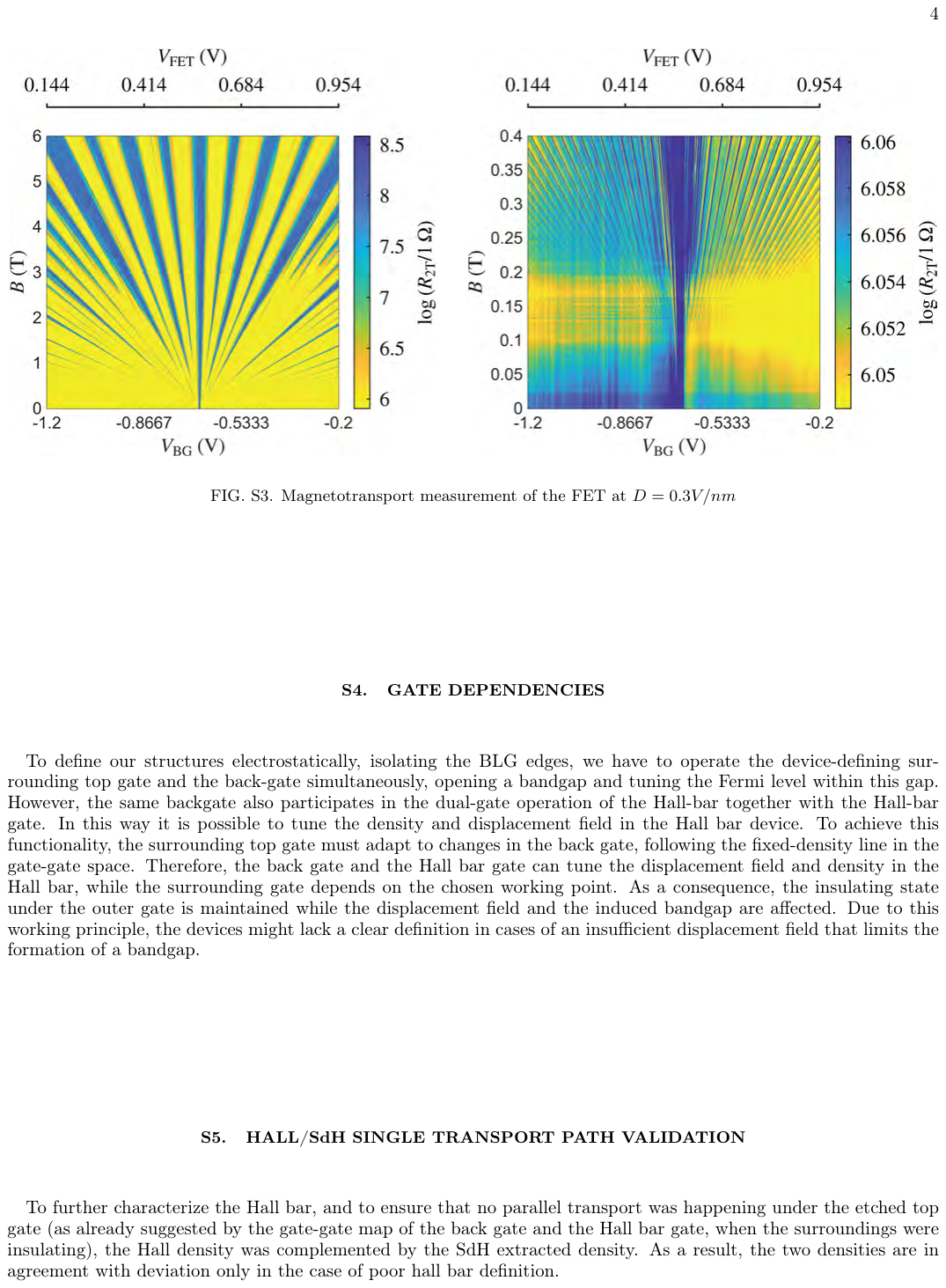}
\end{figure*}

\begin{figure*}[p]
    \centering
    \includegraphics[page=5,width=\textwidth]{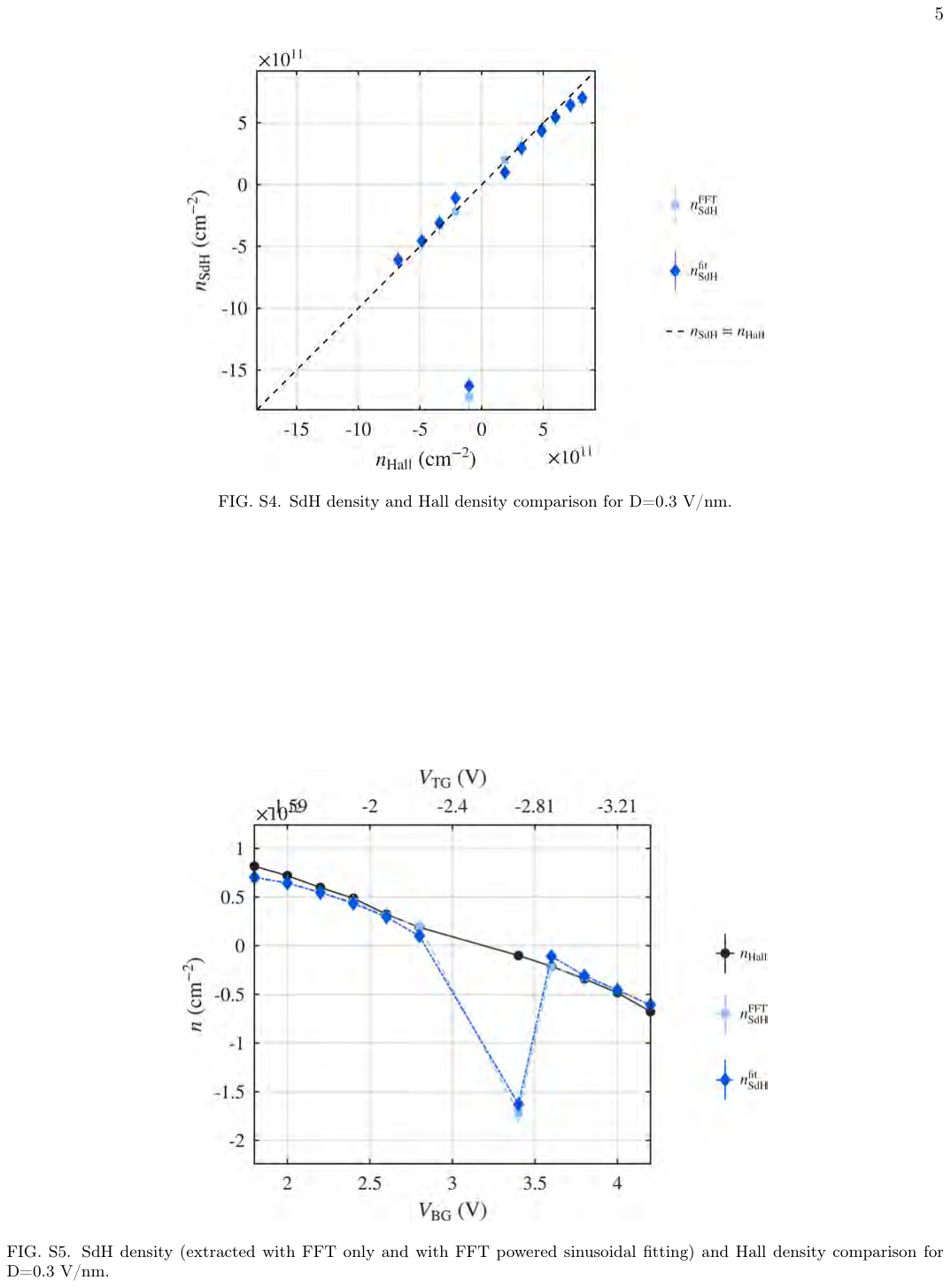}
\end{figure*}

\begin{figure*}[p]
    \centering
    \includegraphics[page=6,width=\textwidth]{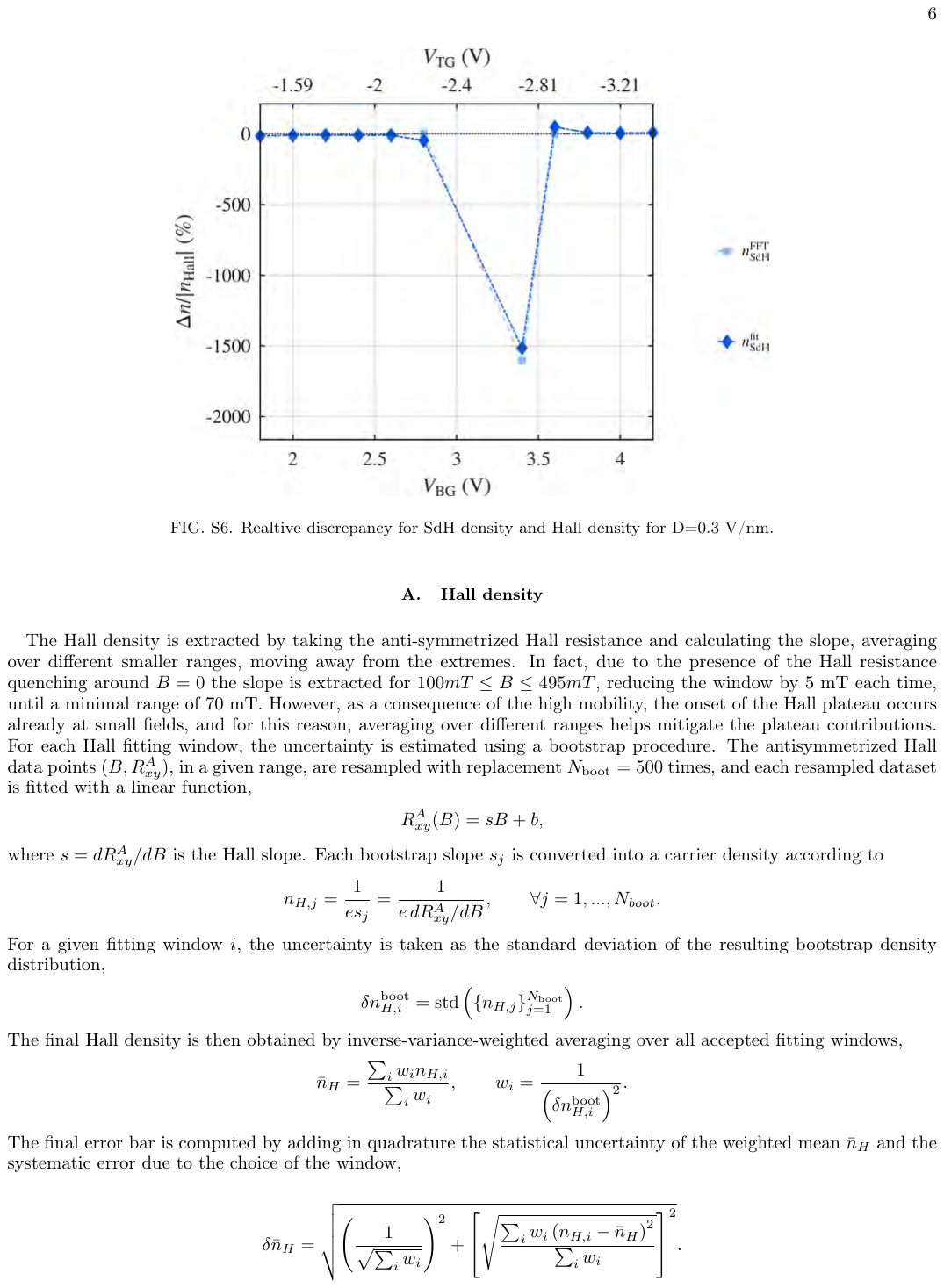}
\end{figure*}

\begin{figure*}[p]
    \centering
    \includegraphics[page=7,width=\textwidth]{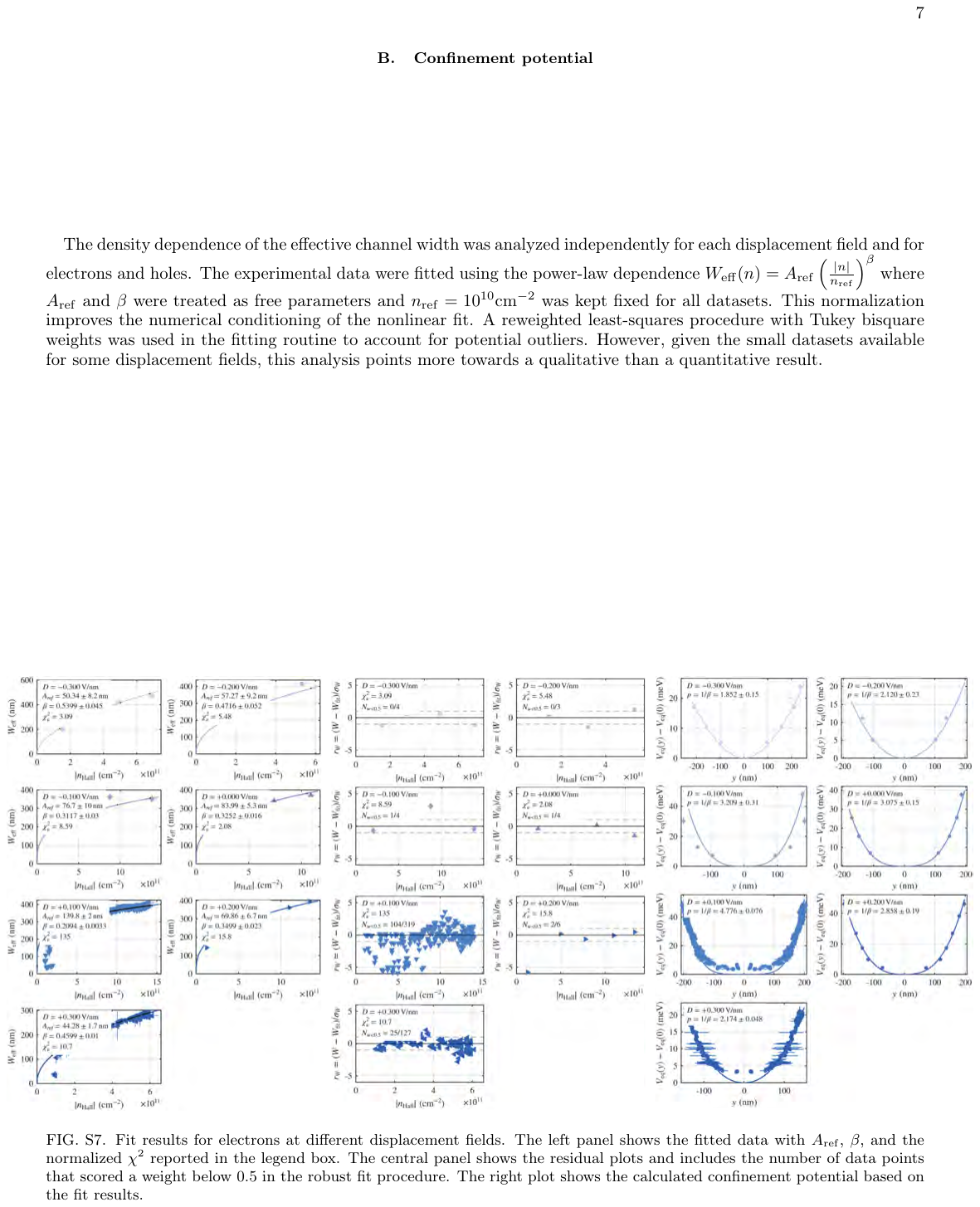}
\end{figure*}

\begin{figure*}[p]
    \centering
    \includegraphics[page=8,width=\textwidth]{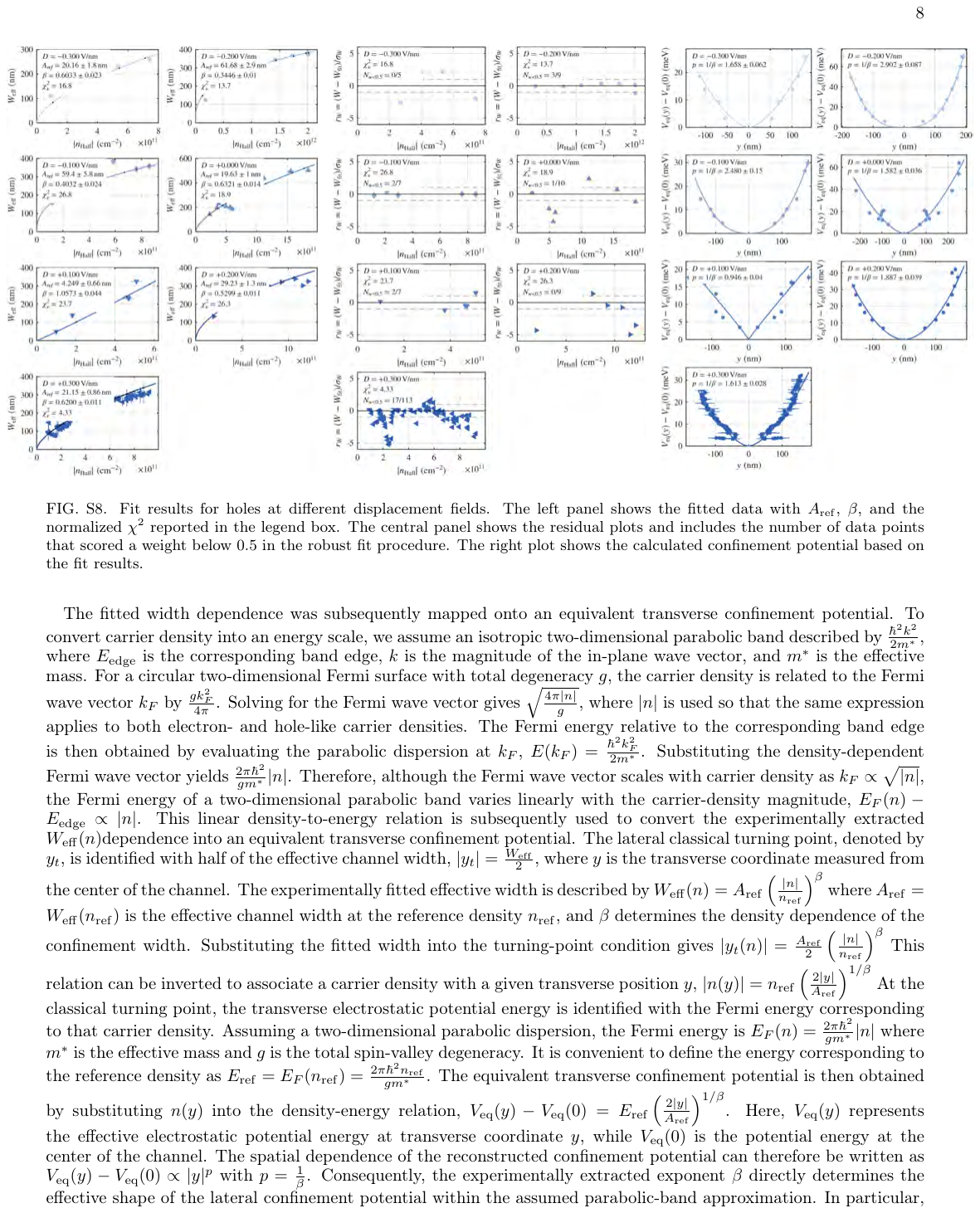}
\end{figure*}

\begin{figure*}[p]
    \centering
    \includegraphics[page=9,width=\textwidth]{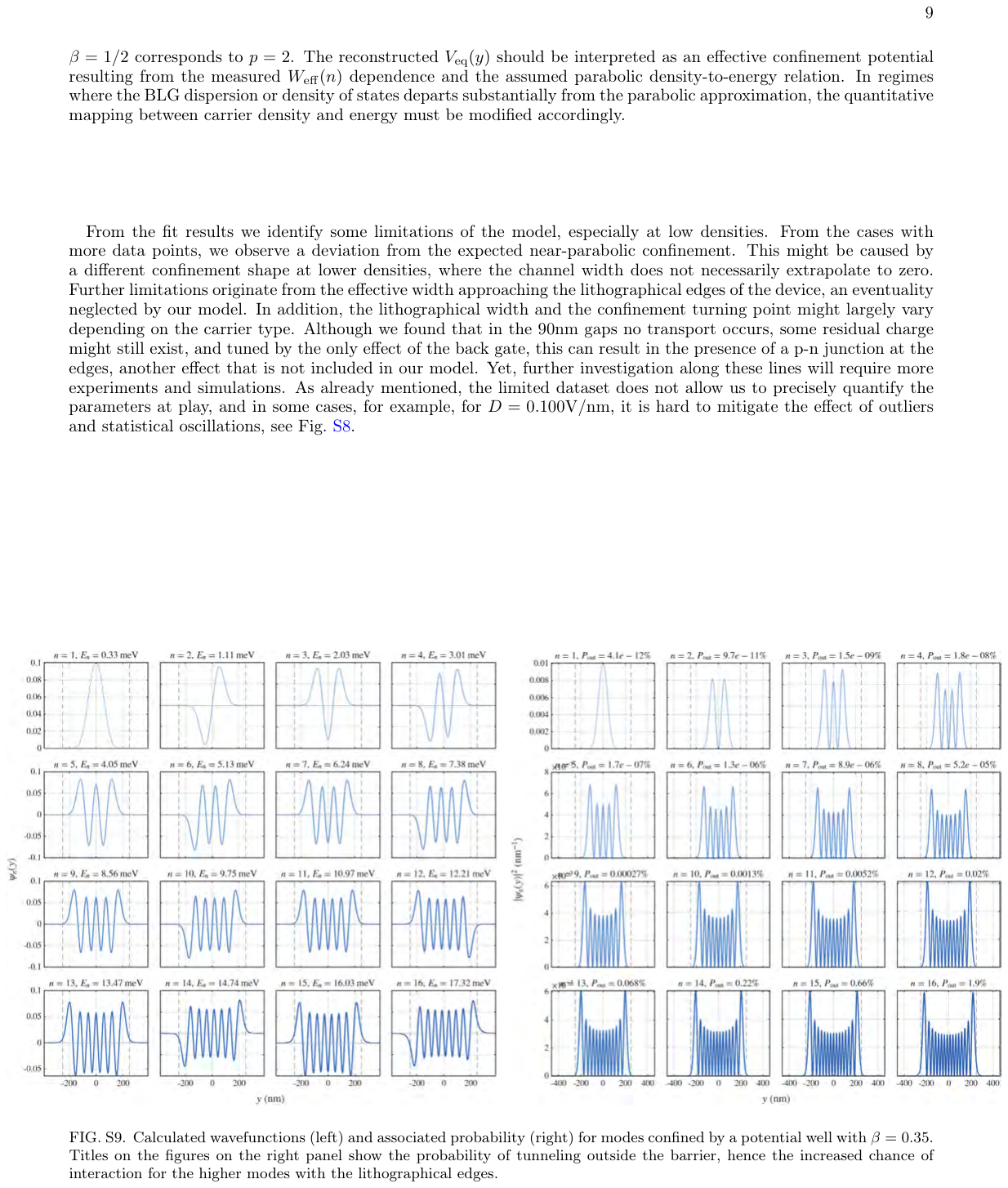}
\end{figure*}

\begin{figure*}[p]
    \centering
    \includegraphics[page=10,width=\textwidth]{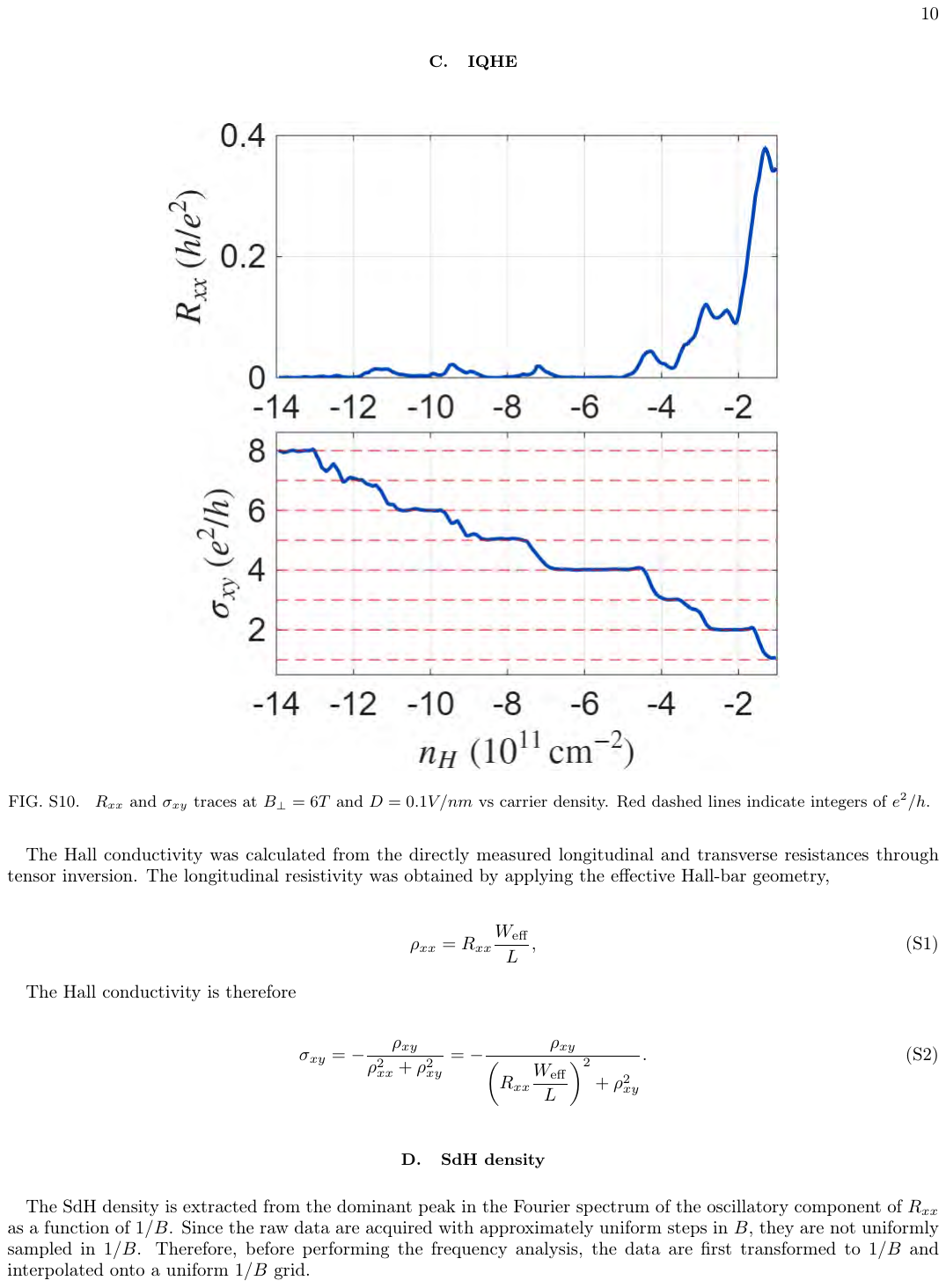}
\end{figure*}

\begin{figure*}[p]
    \centering
    \includegraphics[page=11,width=\textwidth]{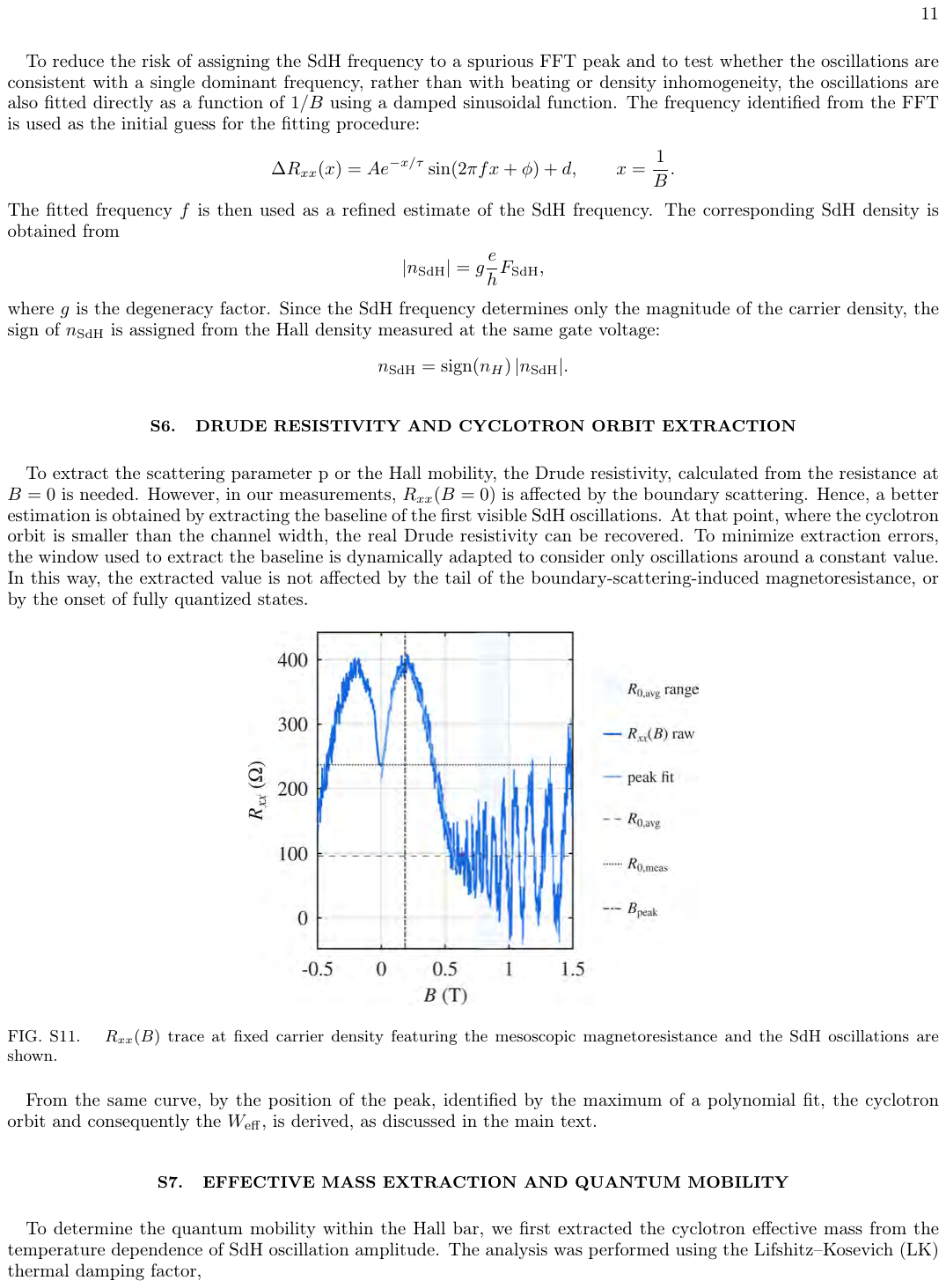}
\end{figure*}

\begin{figure*}[p]
    \centering
    \includegraphics[page=12,width=\textwidth]{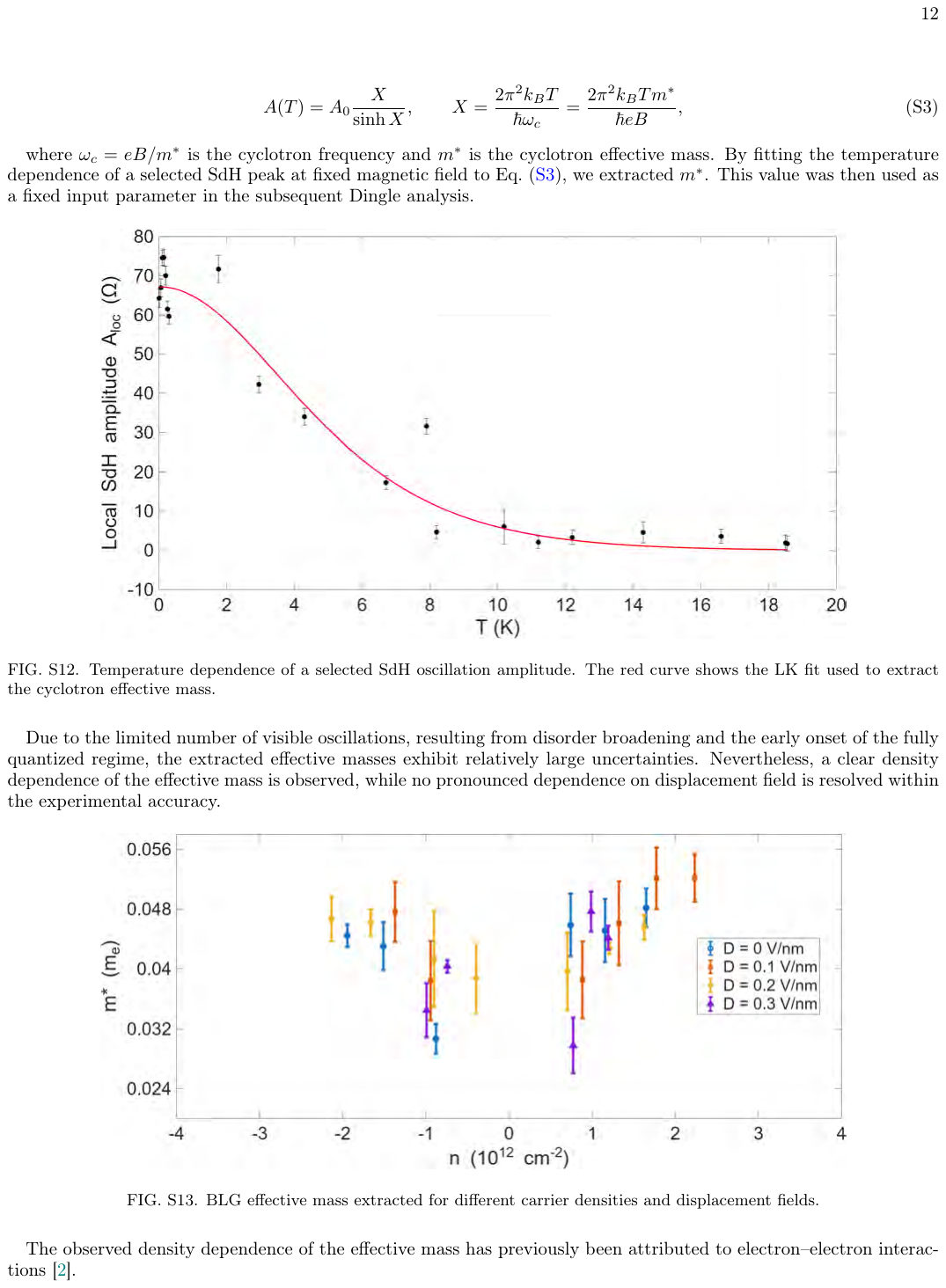}
\end{figure*}

\begin{figure*}[p]
    \centering
    \includegraphics[page=13,width=\textwidth]{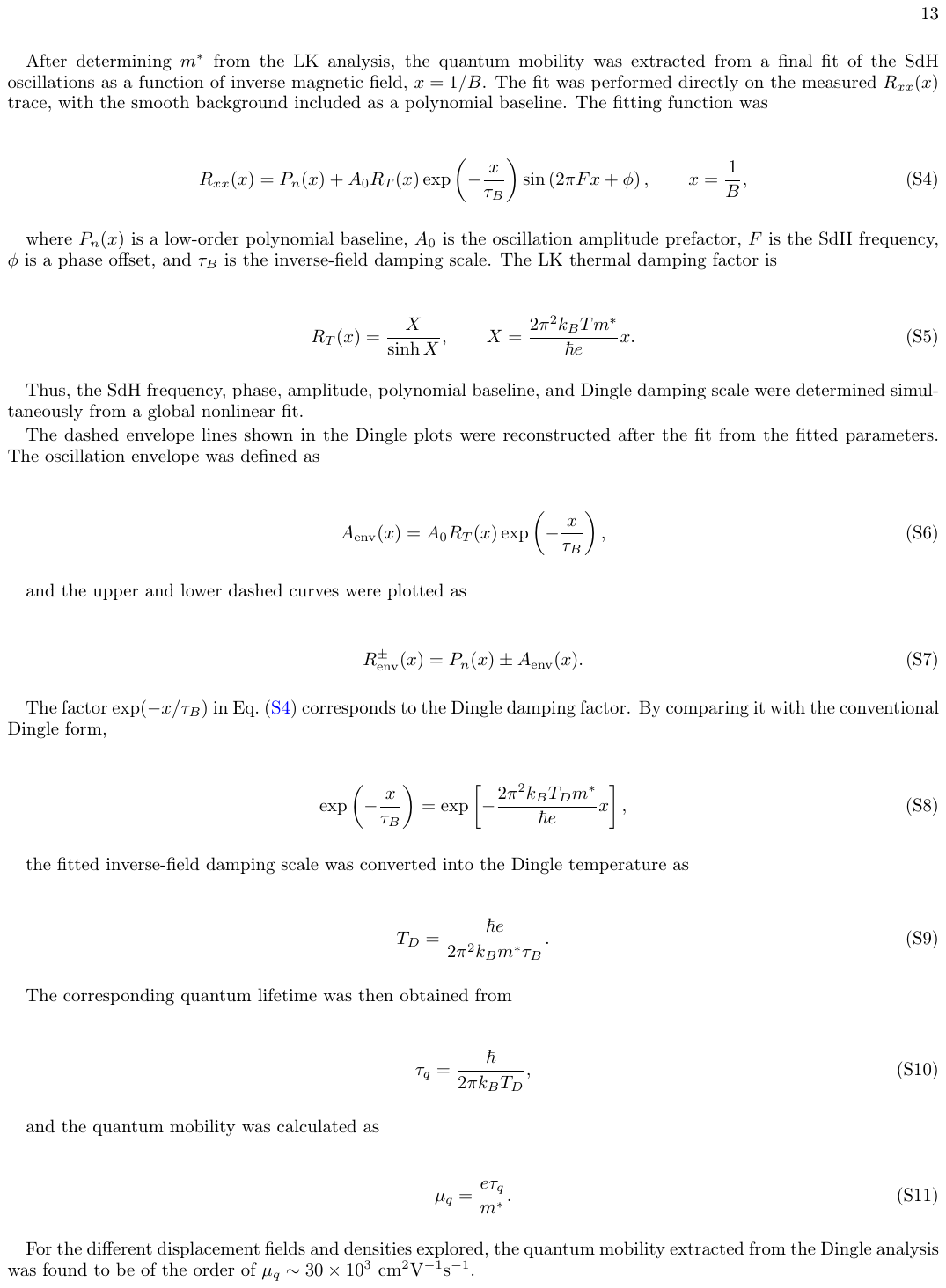}
\end{figure*}

\begin{figure*}[p]
    \centering
    \includegraphics[page=14,width=\textwidth]{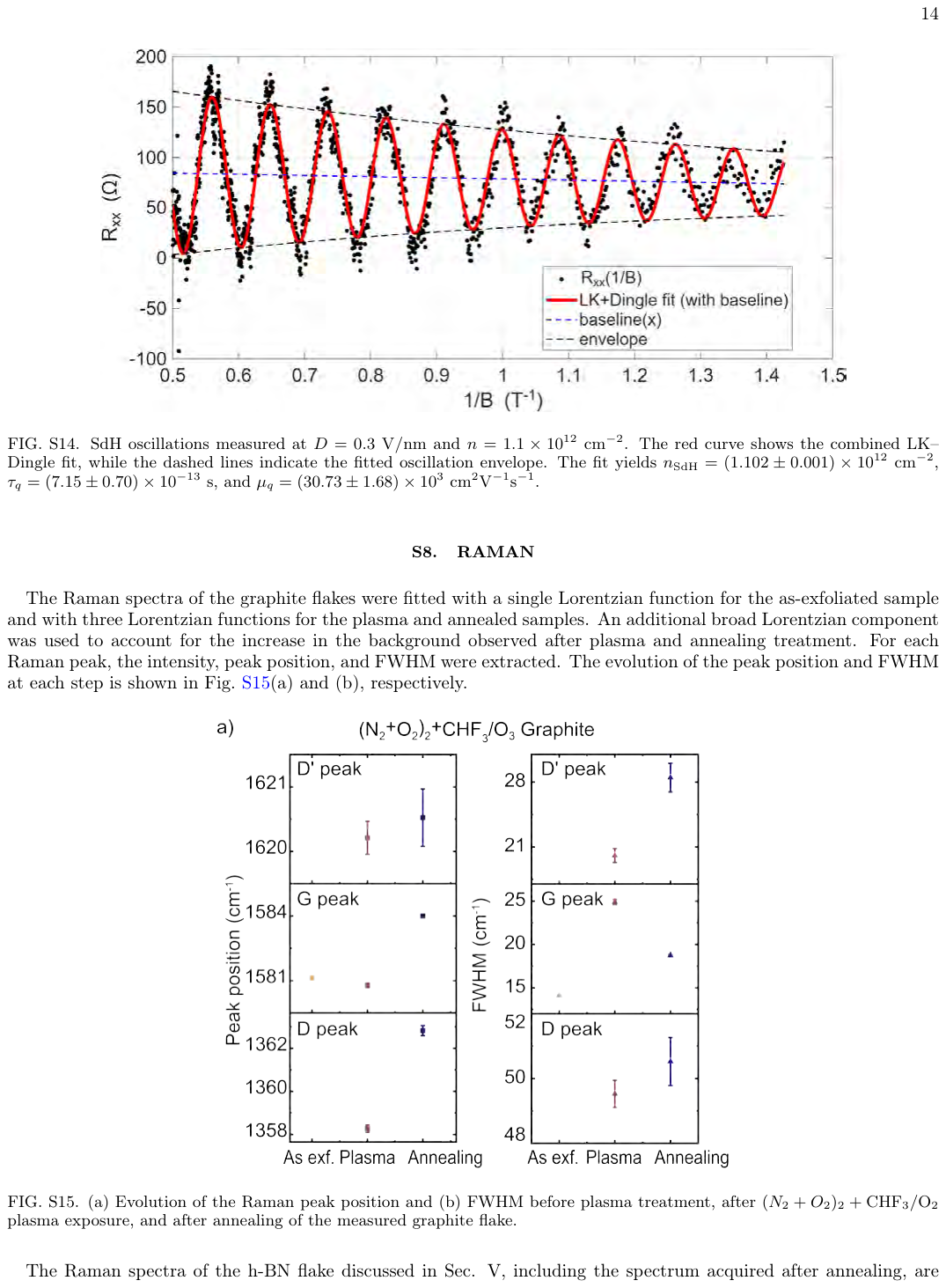}
\end{figure*}

\begin{figure*}[p]
    \centering
    \includegraphics[page=15,width=\textwidth]{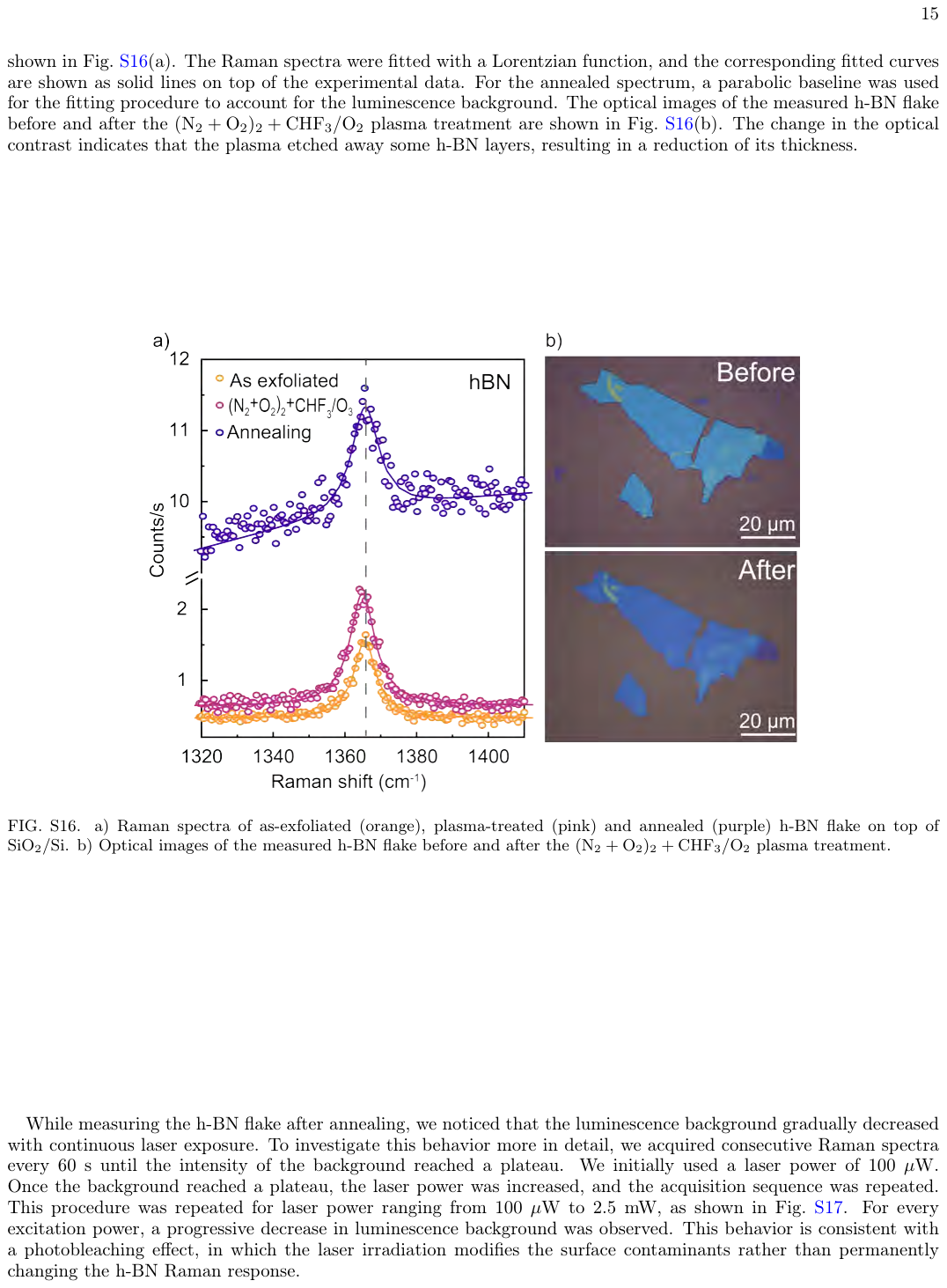}
\end{figure*}

\begin{figure*}[p]
    \centering
    \includegraphics[page=16,width=\textwidth]{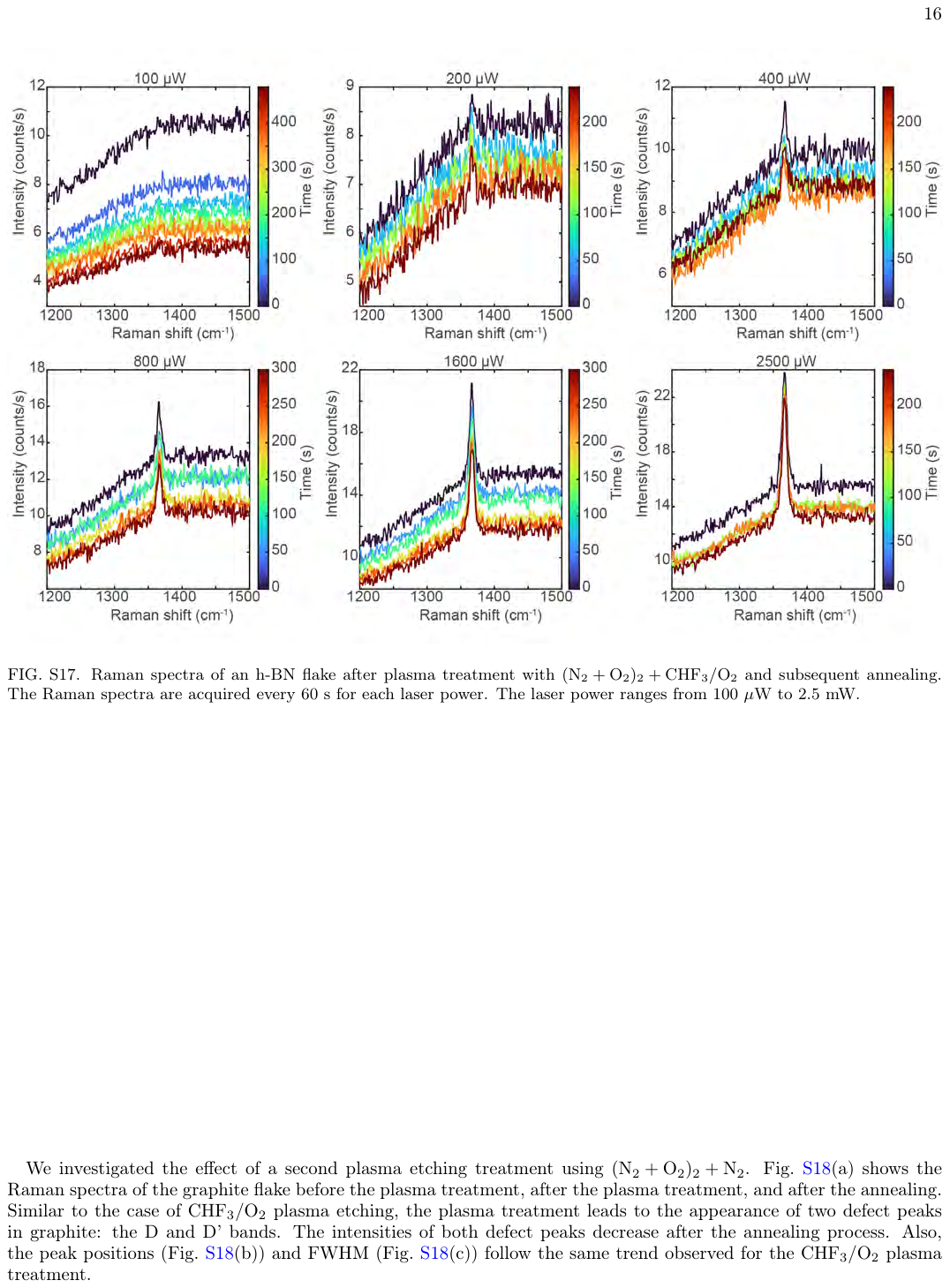}
\end{figure*}

\begin{figure*}[p]
    \centering
    \includegraphics[page=17,width=\textwidth]{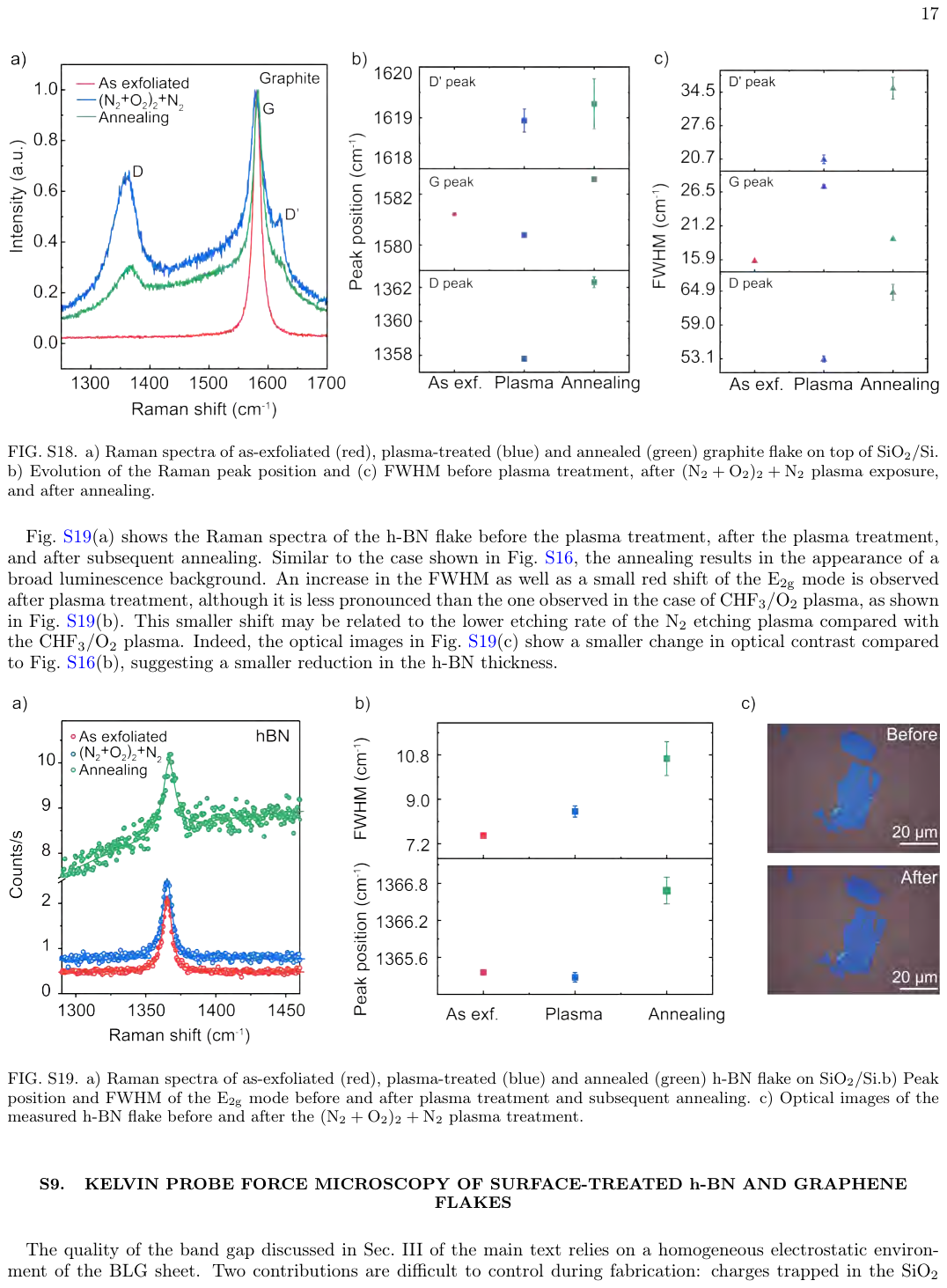}
\end{figure*}

\begin{figure*}[p]
    \centering
    \includegraphics[page=18,width=\textwidth]{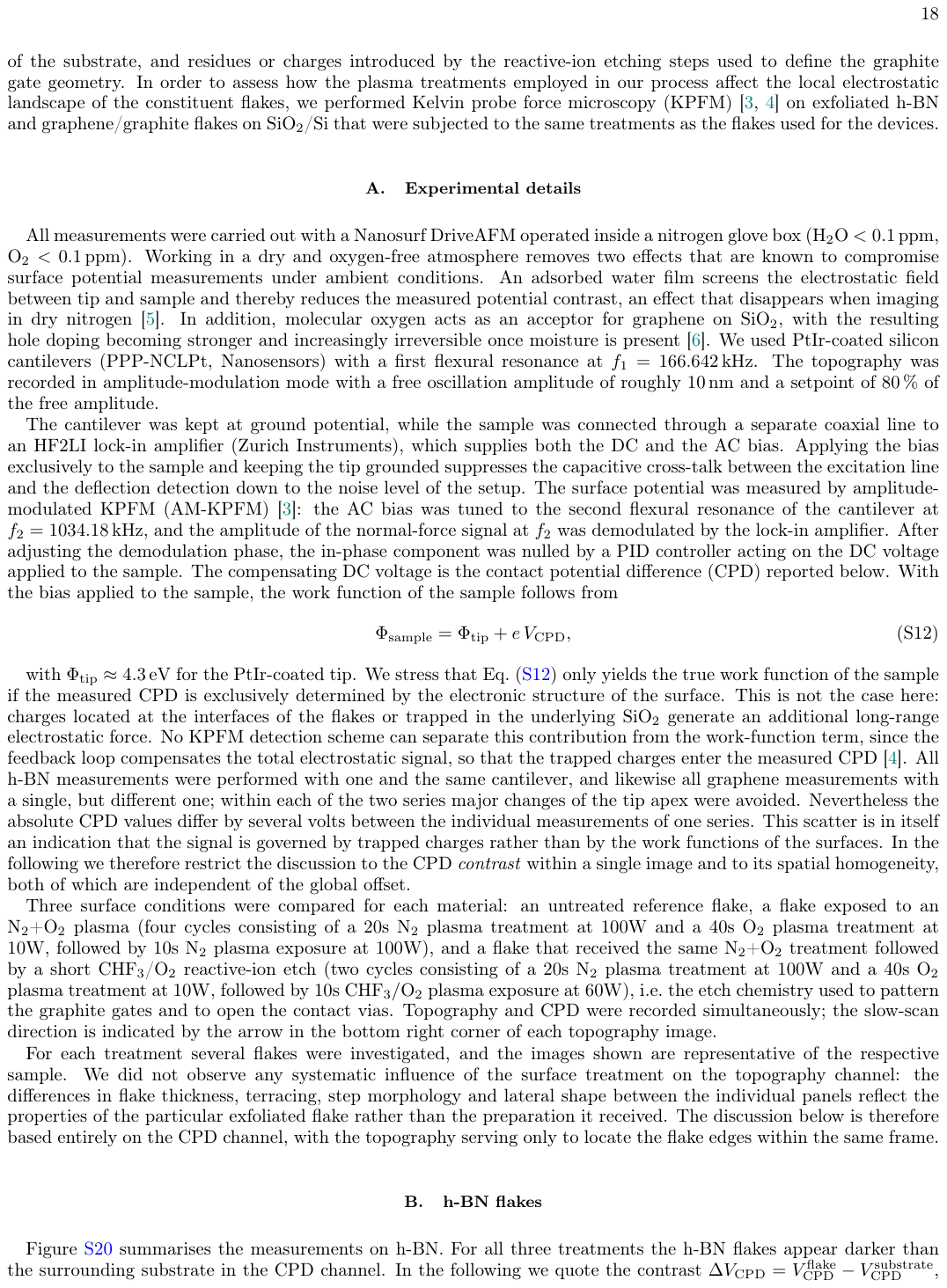}
\end{figure*}

\begin{figure*}[p]
    \centering
    \includegraphics[page=19,width=\textwidth]{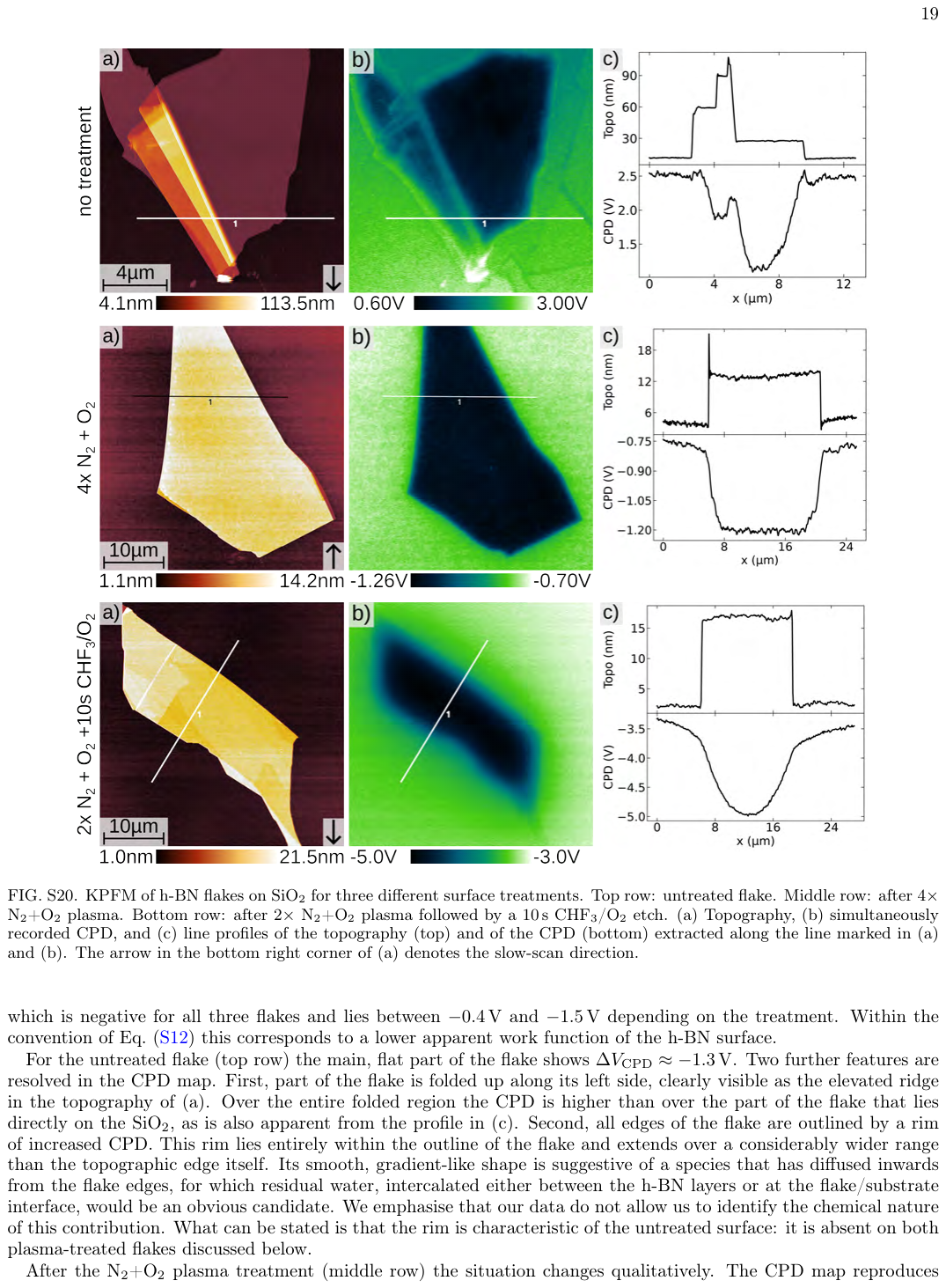}
\end{figure*}

\begin{figure*}[p]
    \centering
    \includegraphics[page=20,width=\textwidth]{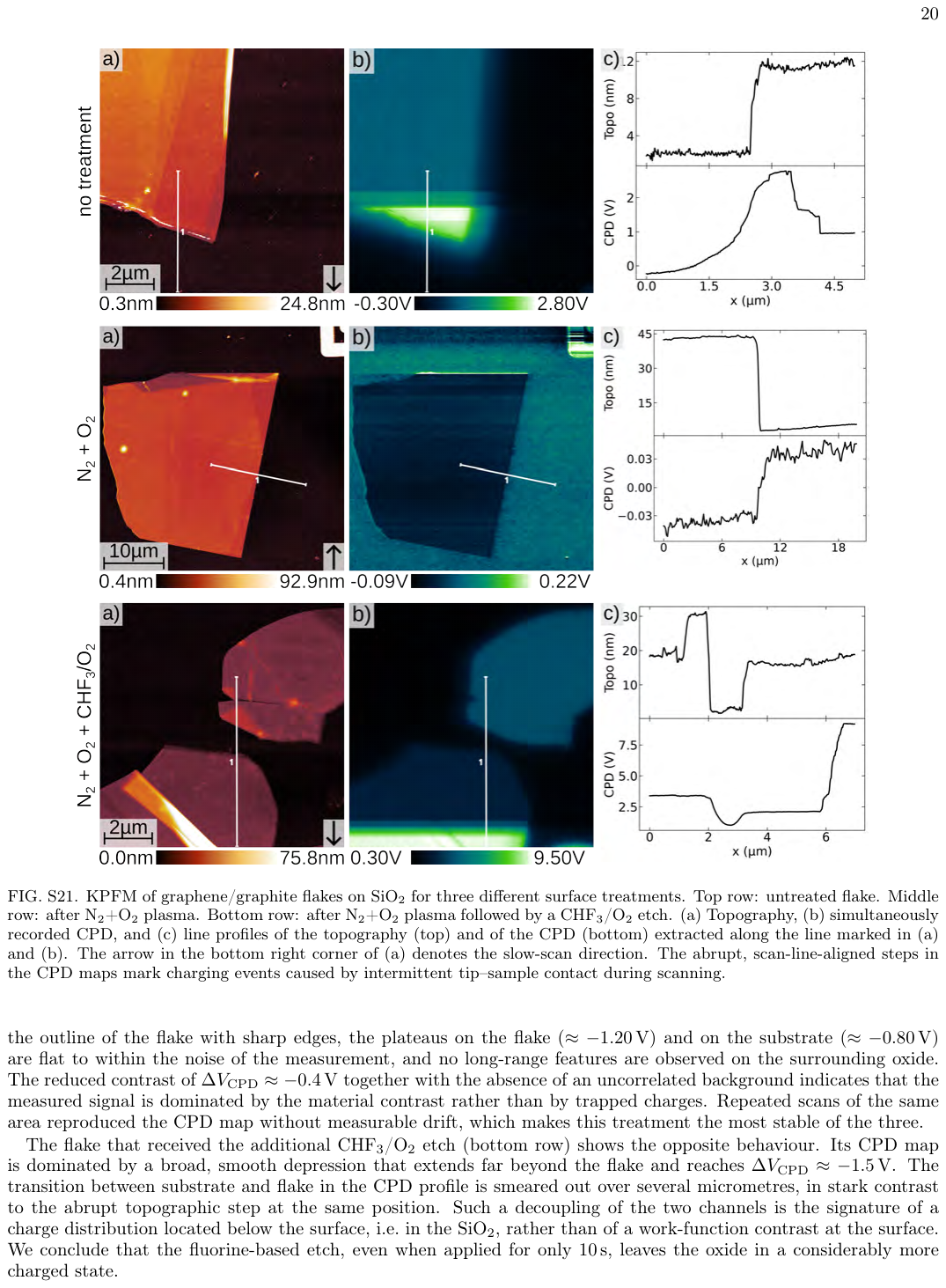}
\end{figure*}

\begin{figure*}[p]
    \centering
    \includegraphics[page=21,width=\textwidth]{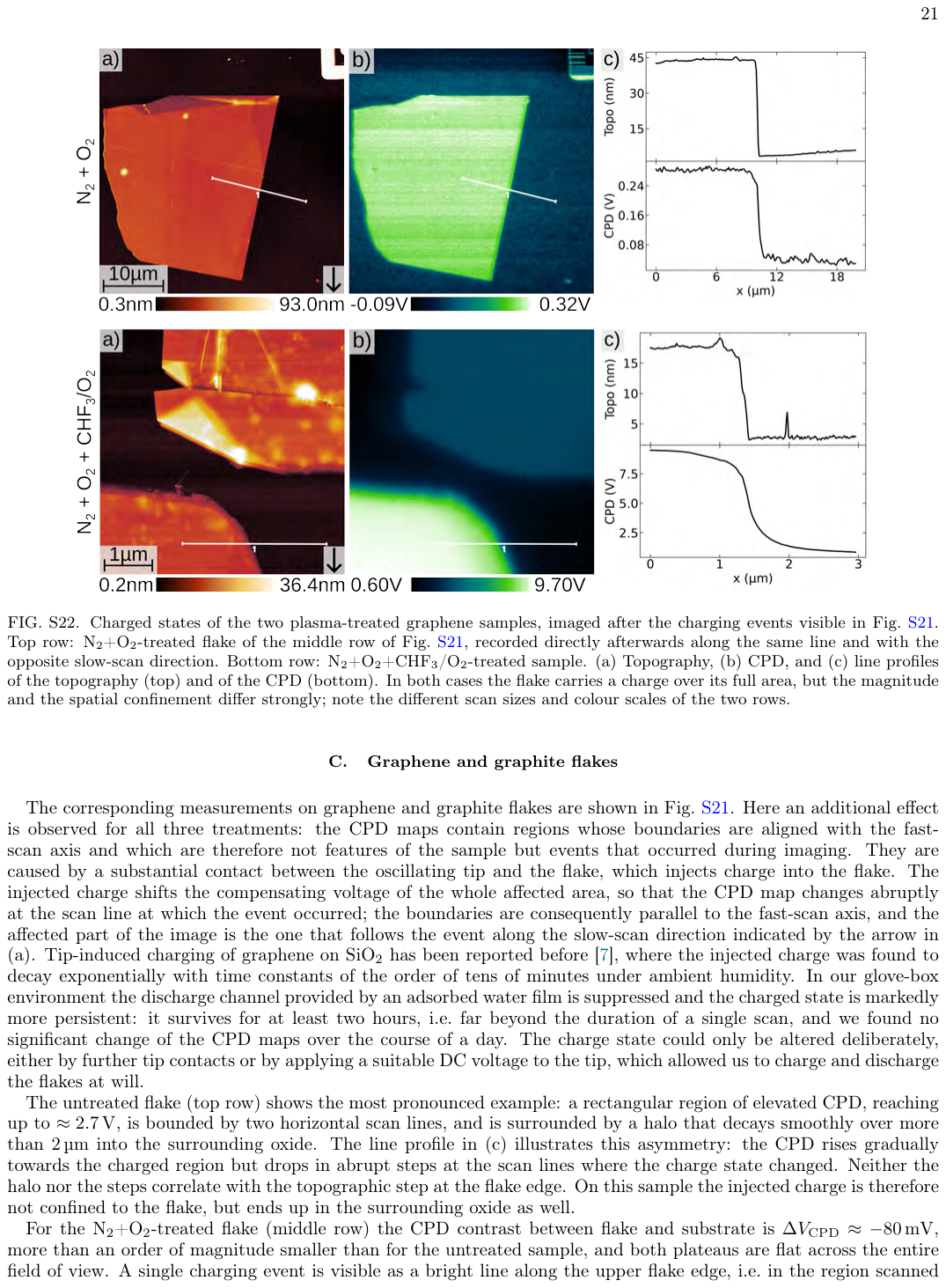}
\end{figure*}

\begin{figure*}[p]
    \centering
    \includegraphics[page=22,width=\textwidth]{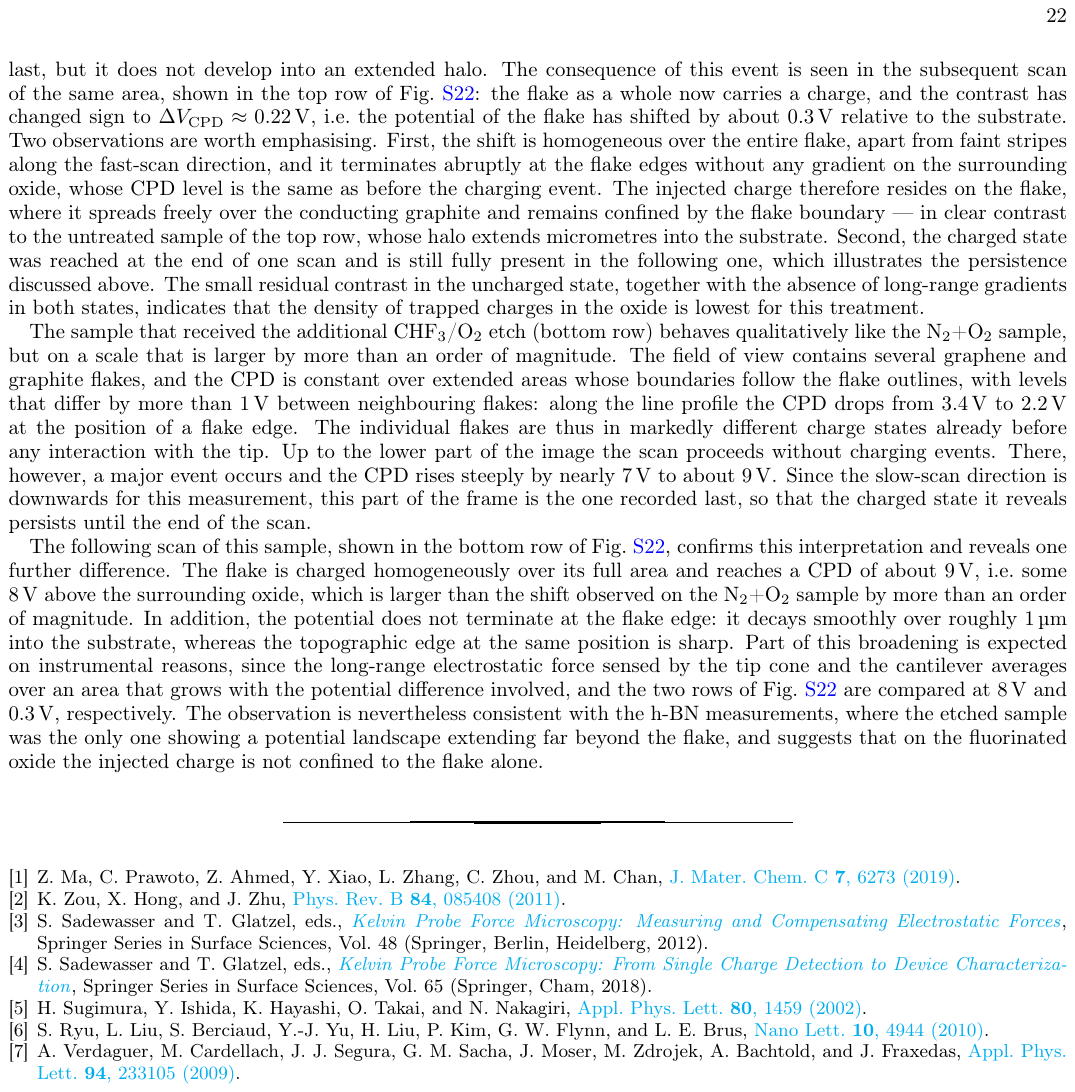}
\end{figure*}

\end{document}